\documentclass[%
 reprint,
superscriptaddress,
 nofootinbib,
 amsmath,amssymb,
 aps,
 prb,
]{revtex4-2}

\usepackage{graphicx}
\usepackage{dcolumn}
\usepackage{bm}
\usepackage[colorlinks=true,linkcolor=blue,citecolor=blue]{hyperref}
\usepackage{verbatim}

\usepackage{xcolor}

\usepackage{braket}

\usepackage{siunitx}

\usepackage{xfrac,faktor}

\usepackage[normalem]{ulem}

\usepackage[capitalise]{cleveref}

\usepackage{verbatim}

\usepackage{multirow}

\usepackage{aas_macros}

\renewcommand{\vector}[1]{\textbf{#1}}

\newcommand{\hubble}{\text{H}_{0}}

\DeclareSIUnit{\msun}{\text{M}_{\odot}}

\DeclareSIUnit{\year}{\text{yr}}

\DeclareSIUnit{\radiant}{\text{rad}}

\DeclareSIUnit{\degree}{\text{deg}}

\DeclareSIUnit{\parsec}{\text{pc}}

\newcommand{\mchirp}{\mathcal{M}}

\newcommand{\de}[1]{\partial_{#1}}

\newcommand{\snr}{S/N}
\newcommand{\SNR}{\biggl( \frac{S}{N} \biggr)}

\newcommand{\psd}{S_{n}}

\newcommand{\inner}[2]{(#1|#2)}

\newcommand{\likelihood}{\mathcal{L}}

\newcommand{\evidence}{\mathcal{Z}}

\newcommand{\prior}{\pi}

\newcommand{\lcdm}{\Lambda \text{CDM}}

\newcommand{\Om}{\Omega_{m}}

\newcommand{\Ol}{\Omega_{\Lambda}}

\newcommand{\dl}{d_{L}}

\newcommand{\FM}{\Gamma}

\newcommand{\CM}{\Sigma}

\newcommand{\epsinv}{\varepsilon_{\text{inv}}}

\newcommand{\cosmoset}{\mathcal{S}}

\begin{document}

\preprint{APS/123-QED}

\title{Multiband gravitational wave cosmology with stellar origin black hole binaries}

\author{Niccol\`o Muttoni}
\email{n.muttoni@campus.unimib.it}
\affiliation{Department of Physics G. Occhialini, University of Milano - Bicocca, Piazza della Scienza 3, 20126 Milano, Italy}
\affiliation{Laboratoire des 2 Infinis - Toulouse (L2IT-IN2P3), Universit\'e de Toulouse, CNRS, UPS, F-31062 Toulouse Cedex 9, France}

\author{Alberto Mangiagli}
\affiliation{Department of Physics G. Occhialini, University of Milano - Bicocca, Piazza della Scienza 3, 20126 Milano, Italy}
\affiliation{APC, AstroParticule et Cosmologie, Université de Paris, CNRS, F-75013 Paris, France}

\author{Alberto Sesana}
\affiliation{Department of Physics G. Occhialini, University of Milano - Bicocca, Piazza della Scienza 3, 20126 Milano, Italy}
\affiliation{INFN, Sezione di Milano-Bicocca, Piazza della Scienza 3, 20126 Milano, Italy}

\author{Danny Laghi}

\affiliation{Dipartimento di Fisica ``Enrico Fermi'', Universit\`a di Pisa, Largo Pontecorvo 3, I-56127 Pisa, Italy}
\affiliation{INFN, Sezione di Pisa, Largo Pontecorvo 3, I-56127 Pisa, Italy}
\affiliation{Laboratoire des 2 Infinis - Toulouse (L2IT-IN2P3), Universit\'e de Toulouse, CNRS, UPS, F-31062 Toulouse Cedex 9, France}

\author{Walter Del Pozzo}

\affiliation{Dipartimento di Fisica ``Enrico Fermi'', Universit\`a di Pisa, Largo Pontecorvo 3, I-56127 Pisa, Italy}
\affiliation{INFN, Sezione di Pisa, Largo Pontecorvo 3, I-56127 Pisa, Italy}

\author{David Izquierdo-Villalba}

\affiliation{Department of Physics G. Occhialini, University of Milano - Bicocca, Piazza della Scienza 3, 20126 Milano, Italy}

\author{Mattia Rosati}
\affiliation{Department of Physics G. Occhialini, University of Milano - Bicocca, Piazza della Scienza 3, 20126 Milano, Italy}

\date{\today}

\begin{abstract}

Massive stellar origin black hole binaries (SBHBs), originating from stars above the pair-instability mass gap, are primary candidates for multiband gravitational wave (GW) observations. Here we study the possibility to use them as effective dark standard sirens to constrain cosmological parameters. The long lasting inspiral signal emitted by these systems is accessible by the future \emph{Laser Interferometer Space Antenna} (LISA), while the late inspiral and merger are eventually detected by third generation ground-based telescopes such as the \emph{Einstein Telescope} (ET). The direct measurement of the luminosity distance and the sky position to the source, together with the inhomogeneous redshift distribution of possible host galaxies, allow us to infer cosmological parameters by probabilistic means. The efficiency of this statistical method relies in high parameter estimation performances. We show that this multiband approach allows a precise determination of the Hubble constant $\hubble$ with just ${\cal O}(10)$ detected sources. For selected SBHB population models, assuming \num{4} (\num{10}) years of LISA observations, we find that $\hubble$ is typically determined at $\sim 2\%$ ($\sim 1.5\%$), whereas $\Om$ is only mildly constrained with a typical precision of \num{30}$\%$ (\num{20}$\%$). We discuss the origin of some outliers in our final estimates and we comment on ways to reduce their presence.

\end{abstract}


\maketitle




\section{Introduction} \label{intro}

Recent analysis, relying on Type Ia supernovae (SNe) \cite{riess2019large} and Planck measurements \cite{aghanim2020planck}, revealed a tension at $4.4\sigma$ on the determination of the Hubble constant, $\hubble$. While the former reported $\hubble = 74.03 \pm 1.42 \, \si{\kilo\meter\per\second\per\mega\parsec}$, the latter derived $\hubble = 67.4 \pm 0.5 \, \si{\kilo\meter\per\second\per\mega\parsec}$. Although a number of possible solutions have been proposed (see, e.g.,~\cite{verde2019tensions}), the tension still persists. \par
The first gravitational wave (GW) detection \cite{abbott2016observation} brightened the future of astrophysical observations. Among the possible theoretical \cite{PhysRevD.103.122002} and cosmological applications \cite{Abbott_2021}, GWs offer a unique opportunity to provide an independent constraint on the $\hubble$ parameter, thus shedding light on the evolution history of our Universe. Coalescing compact object systems are ideal standard sirens in the determination of cosmological distances and their signals bring direct information about the source's luminosity distance. 
However no information is carried about the redshift. If an electro-magnetic (EM) counterpart is detected, the identification of the host galaxy might provide the redshift information necessary to build the $d_L - z$ relation and constrain $\hubble$ \cite{GW170817_H0}. \par
Even if no information about the source redshift is gathered, a measurement of $\hubble$ can be performed by exploiting the statistical properties of the inhomogeneous redshift distribution of possible galaxy hosts \cite{schutz1986determining}. This method has been developed in the last decade \citep[e.g.,][]{holz2005using, 2011ApJ...732...82P,mukherjee2021accurate,del2018stellar} and has been successfully applied to recent LIGO-Virgo observations, although yielding only mild constraints \cite{Abbott_2021}. \par
The future \emph{Laser Interferometer Space Antenna} \citep[LISA,][]{amaro2017laser} will extend the frequency band currently explored by ground based detectors to the \si{\milli\hertz} region. Moreover, the third generation interferometers Einstein Telescope \citep[ET,][]{Punturo2010} and Cosmic Explorer \citep[CE,][]{Reitze2019Cosmic} will remarkably enhance the GW sensitivity from the ground, thus enabling the exploration of a larger portion of Universe. LISA will  detect the early inspiral of stellar black hole binaries (SBHBs) \cite{PhysRevLett.116.231102}. A fraction of these systems will coalesce within only few years from the first LISA observation, becoming observable by ground based detectors, thus fostering a multiband approach. \par
Combining information from ground and space detectors might provide unique scientific outcomes. This is particularly true in the case of multiband SBHB exploitation for cosmological measurements. While LISA will determine the sky position of the GW source to great accuracy, due to the long persistence of the signal in band, ET will pin down its luminosity distance, thanks to the high signal-to-noise ratio ($\snr$). By combining these  two measurements, the origin of the signal can be constrained to a small 3D volume, encompassing only a small number of candidate galaxy hosts. Therefore, multiband GW sources might prove to be a particularly powerful class of ``dark'' standard sirens. \par
The idea of multiband GW sources was first proposed in the context of intermediate mass black holes (IMBHs) \cite{2010ApJ...722.1197A} and later revised in light of the observation of GW150914 \cite{PhysRevLett.116.231102}. Although the idea is appealing, recent estimates of the SBHB merger rate and mass function coupled with the current LISA sensitivity curve result in rather pessimistic prospects of detection. In fact, when considering only BHs below the pair-instability mass gap, LISA might perhaps detect only a handful of such multiband sources \cite{2019PhRvD..99j3004G}. However, things change when the BH mass function is extended beyond the mass gap. As shown in \cite{mangiagli2019merger}, depending on the details of the SBHB population and on the duration of the LISA mission, ``above-gap'' SBHBs might dominate the population of multiband sources \cite{Ezquiaga_2021}, with up to about a hundred of detected systems. \par
Here we exploit SBHBs above the pair-instability gap jointly observed by LISA and ET (which will be our 3G detector default choice) as dark standard sirens to infer the cosmological parameters via statistical methods. We simulate a population of inspiralling SBHBs in LISA, focusing on the systems that are going to coalesce during the LISA time mission. We perform parameter estimation of the multiband binaries under the Fisher matrix formalism with LISA and ground-based detectors. Combining the sky localization uncertainty provided by LISA and the estimate on the luminosity distance by ET, we construct error boxes in the sky. We populate these volumes with realistic galaxy catalogs \cite{izquierdo2020galactic}, and we infer $\hubble$, as well as other cosmological parameters, by applying a statistical nested sampling algorithm. \par
A very similar approach was taken in \cite{del2018stellar}. Their study focused on the inference of cosmological parameters via LISA-only observations of SBHBs below the pair-instability mass gap. Assuming different detector configurations, they found that the Hubble constant is determined between $\sim5\%$ and $\sim2\%$. In the past few years, however, LISA sensitivity has been downgraded by 50\% at the high frequency end \cite{2019CQGra..36j5011R}, and the latest observing run by LIGO/Virgo reported an intrinsic merger rate of $~24 \,\rm Gpc^{-3} yr^{-1}$ for SBHBs \cite{Abbott_2021}. This will inevitably impact the results of \cite{del2018stellar}, yielding a worse estimate of $\hubble$. Therefore, above-gap SBHBs could very well represent the only possibility to perform precision cosmology in this mass range, unless EM counterpart are found to be associated to SBHB mergers \cite{2017ApJ...835..165B}. \par
Even if EM-based measurements are going to be improved by the time LISA and ET start observing, we stress that GW-based observations rely on completely independent assumptions and systematics, and for distance measurements they do not require complex calibrations as for the construction of a cosmic distance ladder. \par
The paper is structured as follows: in \cref{pre} we present the assumptions under which we perform our analyses. In particular, in \cref{binary_pop} we describe the astrophysical GW sources that we consider in our work and their observational properties. In \cref{par_est} we present the analytical framework adopted to perform the SBHB parameter estimation. \Cref{errbox} is devoted to the description of the error box construction and population, and \cref{bayes}, describes the statistical framework and the numerical implementation of the inference of cosmological parameters from SBHB observations. Our main results are presented and extensively discussed in  \cref{results}, while \cref{end} summarizes the main features of our work and lays out future plans.

\section{Preliminaries} \label{pre}

\subsection{Binary population} \label{binary_pop}

A multiband approach requires multiband sources. In this work we explore the realm of SBHBs above the pair-instability mass gap. Current stellar evolution models predict the lack of stellar black holes in the mass range between $\sim$\SI{60}{\msun} and $\sim$\SI{120}{\msun} \cite{spera2017very}. Depending on the metallicity, the hot core of very massive stars might undergo electron-positron pairs production. These processes soften the star's equation of state and bring the star to collapse. The rising temperatures from the contraction then trigger thermonuclear runaway reactions that completely disrupt the progenitor in a luminous \emph{pair-Instability supernova} (PISN) event, thus leaving no remnant \cite{woosley2002evolution}. On the contrary, no physical process can halt the collapse triggered in the last life stages of very massive stars, and the progenitor leaves a BH remnant with mass greater than $\sim$\SI{120}{\msun} \cite{spera2017very}. In this study we will consider SBHBs composed by such ``above-gap'' objects, i.e.~BHs with masses in the range \SI{120}{\msun}-\SI{300}{\msun}, and we will assume the mass gap has a sharp cutoff at $[60,\,  120] \rm M_{\odot}$ (but look also at \cite{2020ApJ...888...76M} for the effects of star rotation and compactness on the BHs mass distribution). \par
When bounded in coalescing binaries, above-gap BHs emit a GW signal which crosses multiple frequency bands: from the \si{\milli\hertz}, where LISA observes the long lasting inspiral phase of the system, up to $\mathcal{O}(10^2)$\si{\hertz}, where ET detects the last-inspiral cycles, the merger, and the ringdown. A multiband signal requires to be detected (i.e.~it has to be revealed with $\snr$ greater than a fixed threshold) by both interferometers, with LISA being the first of them. \par
These loud GW sources were extensively investigated in \cite{mangiagli2019merger} and in this work we rely on those results to perform our analysis. Assuming the optimistic scenario developed by the authors, the LISA merger rate of above-gap GW sources is estimated to be between $R \simeq \SI{10}{\per\year}$ and $R \simeq \SI{14}{\per\year}$. A LISA mission time of \num{4} (\num{10}) years would then reveal $\sim \num{40}$ ($\sim \num{140})$ multiband events.
However, as in \cite{mangiagli2019merger}, we first consider three different subpopulations of SBHBs: below-gap (above-gap) binaries, where both components are below (above) \SI{60}{\msun} (\SI{120}{\msun}), and across-gap binaries with the primary (secondary) BH above (below) \SI{120}{\msun} (\SI{60}{\msun}).
We consider the same models described in that study, i.e.~an optimistic (sSFR-sZ) and pessimistic (mSFR-mZ) model plus two intermediates ones (mSFR-sZ and  sSFR-mZ). The optimistic model features an higher star formation rate and lower metallicity, hence leading to more massive SBHBs than the pessimistic one. We defer to the original paper for extensive details on how the population of SBHBs is built and we summarize here only the main properties of our binary population model. \par
We consider two models for the SFR and the average metallicity evolution as function of redshift. The pessimistic SFR and metallicity are taken both from \cite{Madau_2017} while the optimistic SFR and metallicity are adopted from \cite{Strolger_2004} and \cite{Madau_2014}, respectively. We assume a power-law stellar initial mass function (IMF) $\xi(M_\star, \alpha)\propto M_\star^{-\alpha}$ extending in the range $[8, \, 350] \, \rm M_{\odot}$, with $\alpha = 2.7$ for the pessimistic SFR and $\alpha = 2.35$ for the optimistic one. Single stars are evolved with the code \texttt{SEVN} \cite{spera2017very} and the resulting BHs are combined assuming a flat mass-ratio distribution in $[0.1, 1]$ and a log-flat time delay distribution in [$50 \, \rm Myr$, $t_{\rm Hubble}$]. The frequency distribution in LISA is computed in the quadrupole approximation for circular orbits following \cite{PhysRev.136.B1224}. \par
We stress that our population model is simplified in many ways. Perhaps most importantly, the resulting BH mass function is obtained by evolving individual stars only, neglecting the effect of binary interaction. In fact, binary evolution models have been generally been implemented for stars up to $\approx 150\, \rm M_{\odot}$ \cite{10.1093/mnras/sty1613, 2021MNRAS.508.5028B}, therefore preventing the possibility to study above-gap remnants in binaries. Moreover, we consider only the standard field formation channel, neglecting alternatives such as dynamical capture \cite{2019ApJ...871...91Z}, hierarchical mergers \cite{2021MNRAS.502.2049L} and accretion and mergers in AGN disks \cite{2021ApJ...908..194T}. In this respect, our BH binary population can be considered a ``toy model'' to provide a proof of principle demonstration of the multiband approach. \par
In our analysis there are two important differences with respect to~\cite{mangiagli2019merger}: we normalized the overall rate at $~25 \,\rm Gpc^{-3} yr^{-1}$ to best match the updates from the latest LIGO/Virgo results and we changed the cosmology choosing the same cosmological values used for the generation of the light cones (see \cref{errbox} for more details) for internal consistency. For each sub-population and each model, we generated 30 Monte Carlo realizations of the expected SBHB population. At this step, each binary is characterized by the two component masses, the merging redshift and the initial frequency. The redshift is sampled between $10^{-3}$ and $\sim19$ in log scale, while the initial frequency is in the interval $[\num{2e-4}, 1] \, \rm Hz$. We checked that extending the minimum frequency range to lower frequencies did not affect the number of observable systems. The frequency is then converted into a coalescence timescale $t_c$, via the standard quadrupole approximation \cite{PhysRev.131.435}, assuming circular binaries. Being interested in multiband events, we selected only systems with $t_c < 20 \, \rm years$~\footnote{Even though LISA will be able to detect systems much further from coalescence, these SBHBs will be characterized by large error (especially on luminosity distance) and therefore will add little information to the measurement of cosmological parameters.}. Assuming a LISA mission lifetime of 10 years, we expect exquisite estimates on the coalescence time for those systems, of the order $\Delta t_c \simeq [1 - 10] \, \rm s$ \cite{PhysRevLett.116.231102}, allowing the unequivocal identification of the binary by ground-based detectors.

\subsection{Parameter estimation formalism} \label{par_est}

To simulate observations with LISA and ET, we evaluate the error on the estimated parameters of each source by means of the Fisher information matrix. The general output of a detector is a time series $s(t)$ given by the superposition of the noise contribution $n(t)$ and a GW signal $h(t)$, if present:

\begin{equation}
    s(t) = n(t) + h(t) \, .
\end{equation}

Assuming stationary, Gaussian white noise, the $\snr$ produced by the GW signal is described by

\begin{equation}
    \SNR^2 = \inner{h}{h} \, ,
\end{equation}
where the round brackets $(\,|\,)$ refer to the inner product between two real functions $A(t)$ and $B(t)$ defined as

\begin{equation}
\inner{A}{B} = 4 \, \text{Re} \int_{0}^{+\infty} df \, \frac{\tilde A^{*}(f) \tilde B(f)}{\psd (f)} \, .
\label{eqn:inner}
\end{equation}

Here the tilde labels Fourier transformed quantities, while the star denotes complex conjugate quantities. The $\psd$ term represents the \emph{power spectral density} (PSD) of the detector in \si{\per\hertz} units. \par 
The GW signal $h(t,\Theta)$ produced by a spinning, precessing, and eccentric binary in a detector is  characterized by a set of \num{17} parameters ${\bf{\Theta}}= \{\Theta_{1}, \, ... \,, \Theta_{17}\}$. In the limit of high $\snr$, the probability (or likelihood) that the observed signal is described by $\bf{\Theta}$, given an output $s(t)$, is

\begin{equation}
    p({\bf{\Theta}}|s) \propto \exp \biggl[ -\frac{1}{2}\inner{\de{i}h}{\de{j}h} \Delta\Theta_{i}\Delta\Theta_{j} \biggr] \, ,
    \label{eqn:error_distribution}
\end{equation}
where $\de{i}$ denotes the derivative of the signal $h(t,\Theta)$ with respect to the parameter $\Theta_i$. Equation \eqref{eqn:error_distribution} describes a multivariate Gaussian distribution centered in $\bf{\Theta}$ and with covariance matrix $\Sigma = \inner{\de{i}h}{\de{j}h}^{-1}$. The Fisher information matrix $\FM$ is then defined as

\begin{equation}
    \FM_{ij} = \inner{\de{i}h}{\de{j}h} \, ,
    \label{eqn:FM}
\end{equation}
and the parameter uncertainties and covariances are thus contained in its inverse $\CM$. Moreover, since LISA and ET are independent detectors, we can construct a more informative Fisher matrix by simply adding the individual ones:

\begin{equation}
\label{eqn:LISA_ET}
    \FM_{\text{ET+LISA}} = \FM_{\text{ET}} + \FM_{\text{LISA}} \, .
\end{equation}

The new matrix contains the features of both detectors, therefore yielding better constraints on source parameters, in particular on sky location and luminosity distance. \par
For each binary sampled from our distributions as described in \cref{binary_pop}, we compute the $\snr$ and Fisher matrix in LISA. We adopt the same sensitivity curve described in \cite{PhysRevD.102.084056}. The signal in LISA is described by the inspiral-only precessing waveform presented in \cite{PhysRevD.90.124029}, assuming random spin magnitude in $[0,1]$ aligned with the binary angular momentum for consistency with the waveform adopted for ground-based detectors, as discussed below. Sky position and direction of the binary orbital angular momentum are randomly sampled from a uniform distribution on the sphere. We choose $\snr = 8$ as threshold for LISA detection and for each detected binary we compute the $15 \times 15$ Fisher matrix according to \cref{eqn:FM}. Due to the numerical nature of the problem, we must check the outcome of the subsequent inversion process. We checked the discrepancy between the matrix product $(\FM \cdot \CM)_{ij}$ and the Kronecker symbol $\delta_{ij}$ through

\begin{equation}
    \epsinv = \max_{i, \, j} |(\FM \cdot \CM)_{ij} - \delta_{ij} | \, ,
\label{eqn:epsinv}
\end{equation}
and we consider the inversion successful if $\epsinv \le \num{e-3}$ (see the appendix in \cite{berti2005estimating} for details on the procedure). \par
All binaries detected by LISA are then analyzed with a pipeline for ground-based detectors. The ET Fisher matrix [again defined through \cref{eqn:FM}] is computed using the \texttt{PYTHON} library \texttt{PyCBC} \cite{PyCBC}, which can perform the inner products in \cref{eqn:inner} for a number of waveform approximants. Since ET will also detect the merger and ringdown, we adopt the \emph{PhenomD} waveform \cite{khan2016frequency} to model the full GW signal. Due to its domain of definition, the waveform only allows US to study nonprecessing spin-aligned binaries, as mentioned before. Therefore, of the original \num{17} source parameters we neglect the \num{4} corresponding to the $x$ and $y$ spin components of each BH. Two of the remaining source parameters are eccentricity related. However, GW emission eventually circularizes any eccentric orbit (but see also \cite{2021ApJ...907L..20T} for alternative scenarios): we choose to assume only circular binaries, and we therefore ignore these parameters. The binaries and their signals are evolved through the high frequency band probed by ET, starting from \SI{3}{\hertz}, and we set an $\snr$ threshold of \num{12} for the 3G interferometer detection.

\begin{table}[h]
\centering
\begin{tabular}{ l  c }
\hline
\hline
\textbf{Quantity} & \textbf{Parameter} \\
\hline
Mass 1 & $\ln M_1$ \\

Mass 2 & $\ln M_2$ \\

Luminosity distance & $\ln \dl$ \\

Spin 1 & $\chi_1$ \\

Spin 2 & $\chi_2$ \\

RA  & $\varphi_N$ \\

DEC & $\mu_{N} = \cos \theta_N$ \\

Inclination & $\iota$ \\
\hline
\hline
\end{tabular}
\caption{Parameters that characterize the Fisher matrix. We denote with a capital $N$ subscript the celestial coordinates defined through $\theta_N = \frac{\pi}{2} - \text{DEC}$ and $\varphi_N = \text{RA}$. Furthermore, we take the natural logarithm of $M_1$, $M_2$ and $\dl$ so to deal with relative errors.}
\label{tab:selected_parameters}
\end{table}

Even considering circular binaries with aligned spins, we are still left with a $11\times11$ matrix that keeps track of the two BH masses and dimensionless spins, the luminosity distance, right ascension and declination of the source, the inclination and polarization angles, the merger time and phase. However, we find that in the case of ET the matrix is either ill-conditioned or singular \cite{vallisneri2008use}, and the algorithm struggles with the matrix inversion process~\footnote{To invert the matrices we adopt the LU decomposition.}. Thus, we are forced to select the larger subset of parameters that leads to acceptable $\epsinv$ values. For what concerns the aim of this work, we choose to exclude the polarization and phase angle and the merger time. We do not expect the first two angles to have a strong impact on our parameter estimation and LISA will be anyway able to provide accurate estimates of the coalescence time early during the inspiral phase. In other words, our Fisher matrices are computed for the following subset of 8 parameters: the two BH masses $M_1$ and $M_2$ (with the condition $M_1> M_2$), the  luminosity distance $\dl$, the two BH spins $\chi_1$ and $\chi_2$, the right ascension RA and declination DEC of the binary and the inclination $\iota$~\footnote{Although the ET 11$\times$11 matrix turns out to be ill-conditioned, the LISA one is not. As a sanity check supporting the use of a reduced matrix, we computed the errors on the sky position and luminosity distance for LISA-only observations both for an $11\times11$ and $8\times8$ Fisher matrix. We found comparable results in the two cases, with the latter providing slightly more accurate numbers (by a factor $\approx 1.5$) on average.}. In \cref{tab:selected_parameters} we show how we model the remaining quantities. We therefore reduce the Fisher matrices computed by LISA and ET to the 8 aforementioned parameters, add them, and invert the sum to get the correlation matrix. In this sense, we are not just looking at what each individual detector is able to perform, but we are summing the information matrices from the two detectors as in \cref{eqn:LISA_ET}. The diagonal of the correlation matrix $\CM$ contains the variance of each parameter, while the other entries represent their correlations. The error on $\dl$ can be read directly out of the diagonal elements, whereas the sky position uncertainty area, in units of \si{\square\radiant}, is simply recovered through \cite{lang2006measuring}

\begin{equation}
    \Delta \Omega = 2 \pi \sqrt{\CM_{\varphi_{N} \varphi_{N}}\CM_{\mu_{N} \mu_{N}} - (\CM_{\mu_{N} \varphi_{N}})^2} \, ,
    \label{eqn:sky_loc_error}
\end{equation}
where $\CM_{\mu_{N} \mu_{N}}$ and $\CM_{\varphi_{N} \varphi_{N}}$ correspond to the diagonal elements for the right ascension and declination of the source and $\CM_{\mu_{N} \varphi_{N}}$ is their correlation value.

\subsection{Error box construction and population} \label{errbox}

The uncertainties on the luminosity distance and the sky location allow us to constrain the volume where the GW signal comes from, and galaxies inside it represent possible host candidates. We build the error boxes of the events following the procedure outlined in \cite{laghi2021gravitational}. For a given cosmology, a $\dl \pm \sigma_{\dl}$ measure translates in a $z \pm \sigma_z$ interval, which is obtained by inverting the relation

\begin{equation}
    \dl(z) = c \, (1 + z) \int_{0}^{z} \frac{dz'}{\text{H}(z')} \, .
    \label{eqn:dl_z}
\end{equation}

Here $\text{H}(z)$ represents the Hubble parameter as a function of redshift and its expression in a $\lcdm$ Universe is given by

\begin{equation}
    \text{H}(z) = \hubble \sqrt{\Om(1+z)^3 + \Ol} \, .   
\end{equation}

Since each set of cosmological parameters yields a different $d_L-z$ relation, the $z \pm \sigma_z$ redshift interval is extended in order to take into account for the prior ranges of the cosmological parameters one is willing to infer. From a measure of luminosity distance, we therefore construct a redshift range $[z^{-}, \, z^{+}]$, whose bounds represent the lowest and highest $z$ obtained when the cosmology varies within the prior range of the cosmological parameters, which will be specified in \cref{bayes}. Furthermore, due to the peculiar velocity $v_p$ of galaxies, the \emph{cosmological} and \emph{apparent} redshifts of a GW host might differ from each other. This is indeed an uncertainty source that we must account for. We characterize this error with $\sigma_{pv}(z) = (1 + z)\sigma_{v_{p}}/c$, with $\sigma_{v_p} = \SI{500}{\kilo\meter\per\second}$, which is consistent with the standard deviation of the radial peculiar velocity distribution observed in the Millennium run \cite{springel2005simulations}. In conclusion, each $\dl \pm \sigma_{\dl}$ measure translates in a [$z^{-} - \sigma_{pv}(z^{-}), \, z^{+} + \sigma_{pv}(z^{+})$] redshift range. This corresponds to the redshift shell of the Universe encompassing all galaxies with a redshift consistent with the GW measured luminosity distance, once the cosmological prior and peculiar velocities are taken into account. \par
\begin{figure}[b]
    \centering
    \includegraphics[width=0.5\textwidth]{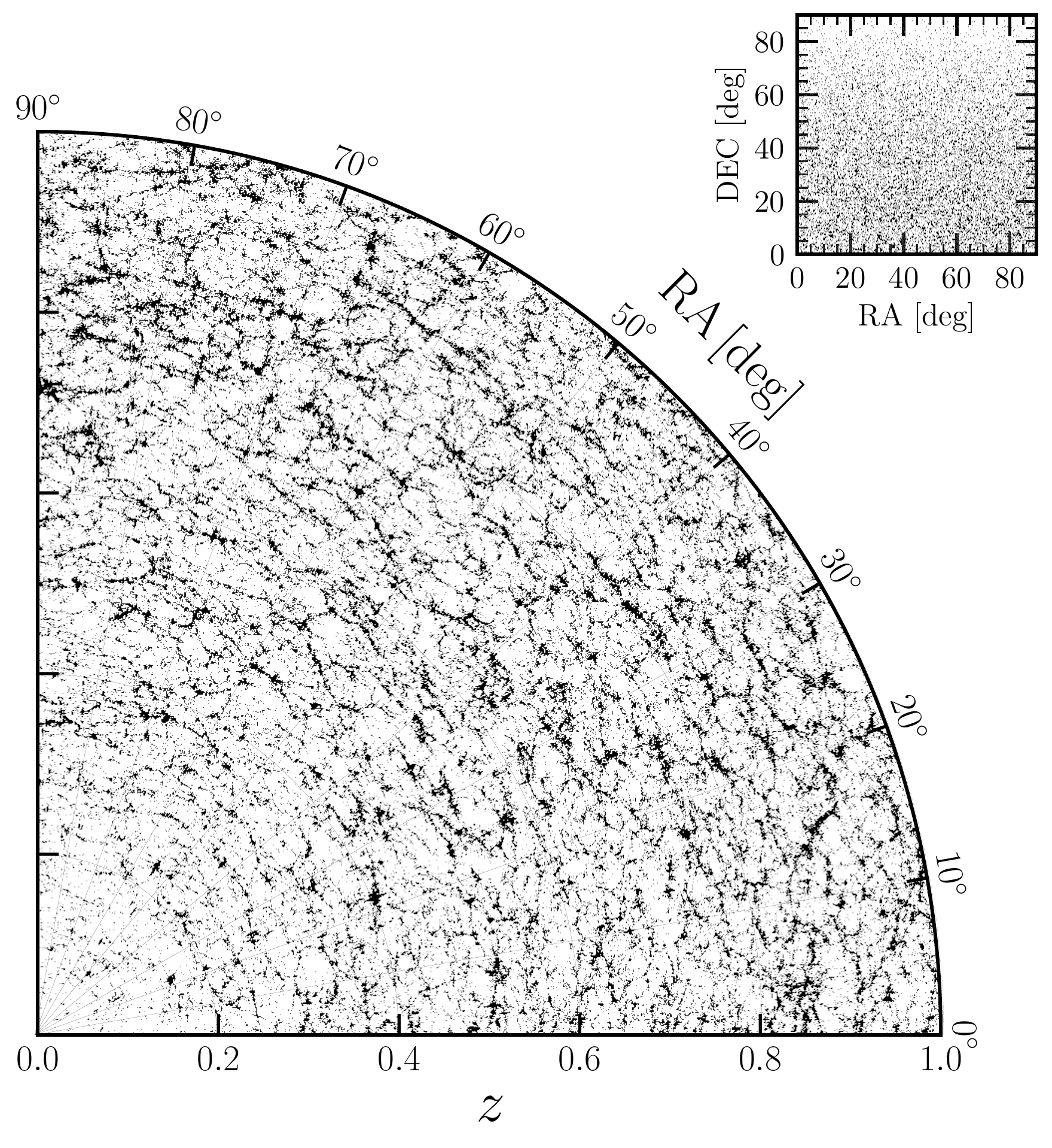}
    \caption{Visual representation of a slice of the light cone adopted in this work. The major panel shows the galaxy distribution in the $z$-RA plane, while the top-right corner plot displays the galaxy distribution in the RA-DEC plane.}
    \label{fig:light_cone}
\end{figure}

The next step is to populate those redshift shells with realistic galaxy catalogs. To this end, we make use of a custom light cone built specifically for this purpose by the authors. In particular, the light cone is assembled based on the methodology presented in \cite{izquierdo2020galactic}, which uses the semianalytical model \texttt{L-Galaxies} applied on the Millennium dark matter merger trees. Specifically, the light cone generated for this work is complete up to $z = 1$ and it contains several physical properties such as mass, magnitudes, observed and geometrical redshift for all the galaxies included in it. Given that we are mainly interested in the low redshift Universe, where our results might be affected by cosmic variance caused by narrow angular apertures, we decide to set the light cone aperture to \num{1}/{8}th of the full sky. We refer the reader to \cite{izquierdo2020galactic} for further details about the light cone construction. As an illustrative example, \cref{fig:light_cone} displays the spatial distribution of the $M_{\rm gal} > \SI{e10}{\msun}$ galaxies~\footnote{The mass resolution of the simulation is $M_{*} = \SI{e10}{\msun}$.} inside ${\sim}\SI{1}{\degree}$ declination slice.

We fix the parameters of the simulation so that $\hubble = \SI{73}{\kilo\meter\per\second\per\mega\parsec}$, $\Om = 0.25$ and $\Ol = 0.75$. Even though these values do not reflect the state-of-the-art estimates, this has no impact on the analysis. Our work aims to show the potential of this GW based measurement for a given Universe, which we are allowed to customize within our pipeline. \par
To locate and populate the error boxes, for each GW event we adopt the following procedure:

\begin{enumerate}
    
    \item We list all the galaxies within our light cone with a cosmological redshift within the $z \pm \sigma_z$ interval consistent with the ET+LISA $\dl \pm \sigma_{\dl}$ measurement and the true (i.e.~the Millennium) cosmology.
    
    \item We randomly~\footnote{However, we reject those selected near the boundary of the light cone, so that we avoid cutting the edges of the $\Delta\Omega$ ellipses.} select a galaxy among them and we denote it as the ``true host'' of the GW event. This galaxy position is described by the set ${\bf{\Theta}}^{\text{th}}=\{d_{L}^{\text{th}}, \mu_{N}^{\text{th}}, \varphi_{N}^{\text{th}}\}$.
    
    \item We draw a $(\mu_N', \varphi_N')$ pair of celestial coordinates from the uncertainty distribution in \cref{eqn:error_distribution} centered in ${\bf{\Theta}}^{\text{th}}$ .
    
    \item We consider a 3$\sigma$ region in $\Delta\Omega$, computed according to \cref{eqn:sky_loc_error} and centered in $(\mu_N', \varphi_N')$. This procedure ensures that the volume is not artificially centered onto the true host, but in a nearby point in the sky consistent with the sky location error given by the GW measurement.
    
    \item We select all the galaxies with $M_{\rm gal} \ge \SI{3e10}{\msun}\footnote{See \cref{gal_mass_threshold} for a discussion on the impact of the adopted mass threshold.}$ within $\Delta\Omega \times [z^{-} - \sigma_{pv}(z^{-}), \, z^{+} + \sigma_{pv}(z^{+})]$ and we associate to each one a hosting probability consistent with the marginalized sky location error given by the GW measurement (see \cref{bayes}). These represent all the possible host candidates for the GW event.
    
\end{enumerate}

\begin{figure*}
    \centering
    \includegraphics[width=0.3\textwidth]{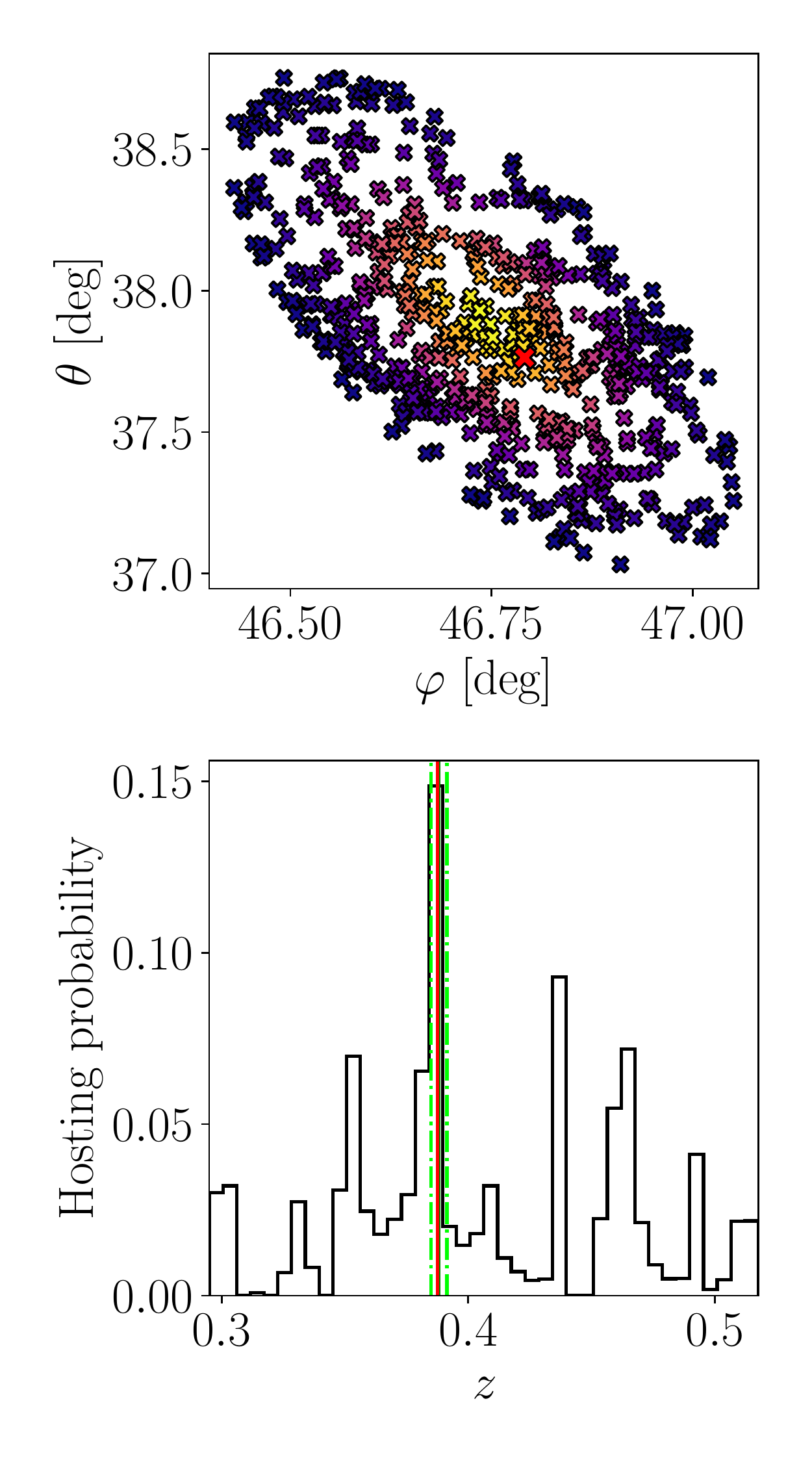}
    \includegraphics[width=0.3\textwidth]{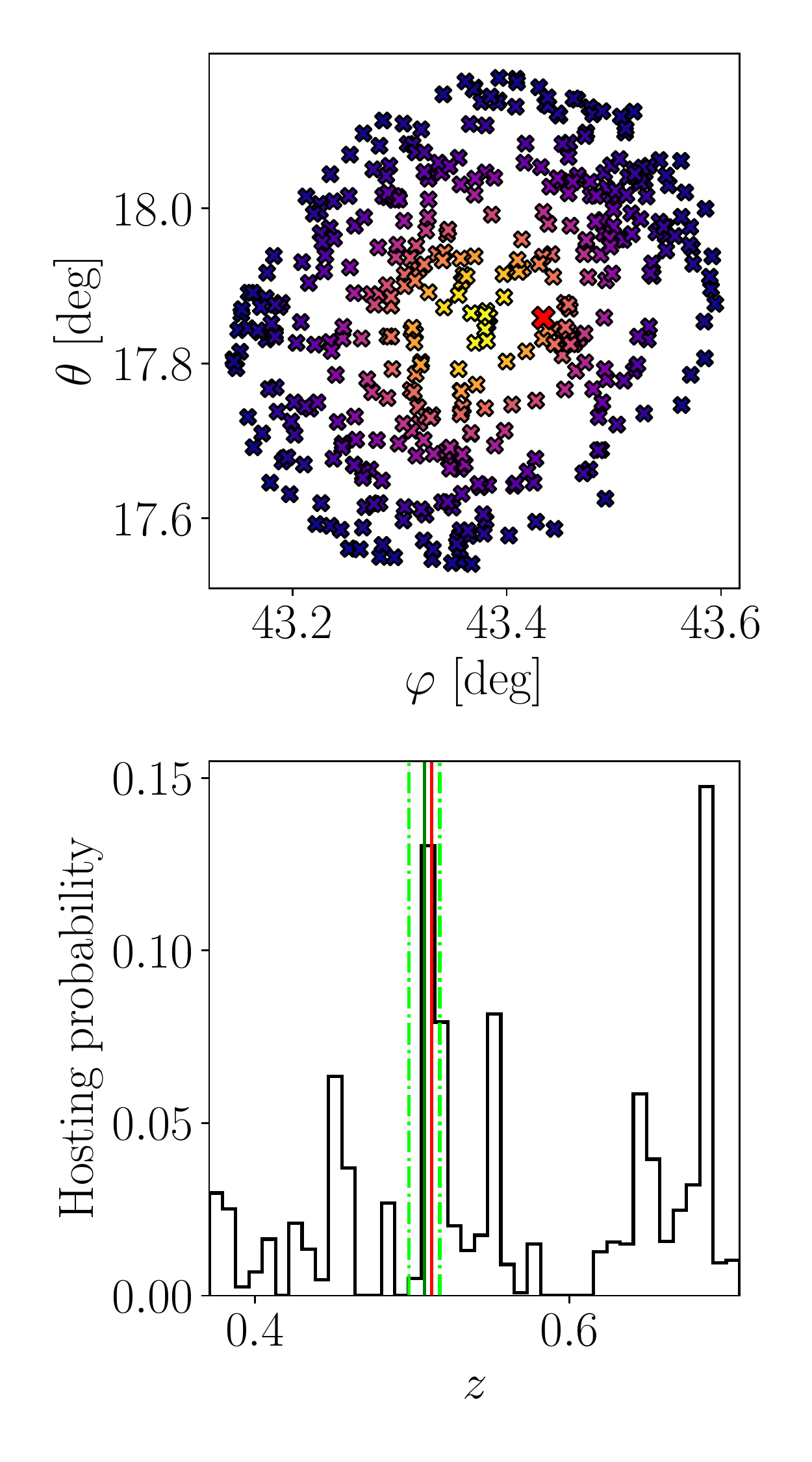}
    \includegraphics[width=0.3\textwidth]{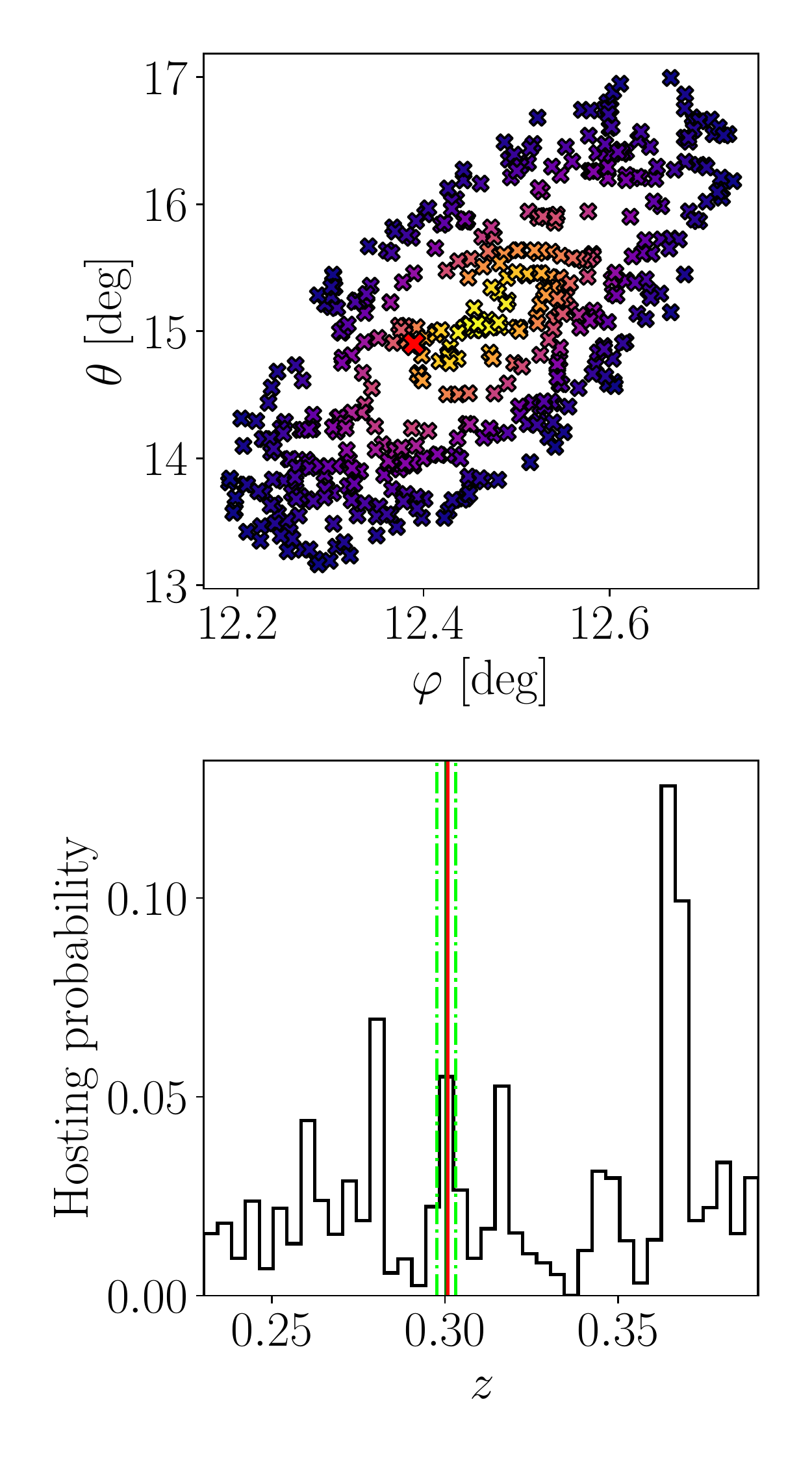}
    \caption{Three different error boxes are displayed column-wise. The top row depicts the distribution of galaxies in the $(\theta,\varphi)$ plane, and the color scale denotes the associated hosting probability, from dark blue (low) to yellow (high). The histograms on the bottom row represent the galaxy distributions over redshift, weighted on the hosting probability. According to the true cosmology, the dark green solid line denotes the best ET+LISA $z$ measurement, while the light green dotted lines represent the $2\sigma$ redshift interval. The red color in each panel denotes the selected ``true host''.
    }
    \label{fig:error_boxes}
\end{figure*}

The outcome of such a method is displayed in \cref{fig:error_boxes}, where we present different error boxes to show the main features and how they may affect the inference of cosmological parameters. In particular, the left column shows an error box with an optimal galaxy clustering in the true cosmology region, thus helping the inference of cosmological parameters. The right column, instead, depicts a noninformative event, due to the misleading information provided by the large galaxy cluster at higher redshift. The middle column is an average error box and it becomes useful to the inference when it is cross correlated with other GW events.

\subsection{Bayesian inference} \label{bayes}

The problem of measuring quantities in nonrepeatable experiments has a Bayesian nature by definition. The entire framework relies on the Bayes theorem, which states that, given a set of data $D$ and a model hypothesis $H$, the probability of $H$ given $D$ is

\begin{equation}
    p(H|D) = \frac{\likelihood(D|H) \prior(H)}{\evidence} \, ,
    \label{eqn:bayes}
\end{equation}
where

\begin{itemize}

    \item $p(H|D)$ is the \emph{posterior distribution}, stating the degree of belief in $H$ after the measure.
    
    \item $\likelihood(D|H)$ is the \emph{likelihood function}, which is known once we assume a model hypothesis.
    
    \item $\prior(H)$ is the \emph{prior distribution}, which amounts to our degree of belief in $H$ before the measure.
    
    \item $\evidence$ is the \emph{evidence}, a central quantity in model selection studies. Since here we are interested just at the posterior distribution, we can neglect this factor and simply renormalize the posterior at the end of the computation.
    
\end{itemize}

In our case, $D$ represents the detected GW events, while $H$ denotes a particular assumed cosmological model that will define the parameter space to be explored. The aim of Bayesian inference is to determine the posterior distribution. To sample the posterior distribution $p(H|D)$ we first need to define the model (which defines the likelihood function $\likelihood$ ) and the prior distribution.

\subsubsection{Single GW event likelihood} \label{single_GW_likelihood}

In the next derivation, we follow the arguments detailed in \cite{del2018stellar}. Consider a set of $n$ GW events $\vector{gw} = \{gw_1, \, ... \, , gw_n\}$ and let $\cosmoset = \{\hubble, \Om, \, ... \}$ be a set of cosmological parameters to be inferred. Each GW event is reasonably independent from the others, therefore the likelihood in \cref{eqn:bayes} can be rewritten as the product of the single GW event likelihoods~\footnote{\label{txt:quasilikelihood}This quantity is more rigorously called \emph{quasilikelihood}, due to the fact that it is obtained through the marginalization of the likelihood over \emph{nuisance} parameters.}:

\begin{equation}
    \likelihood(\vector{gw}|\cosmoset) = \prod_{i = 1}^{n} \likelihood(gw_i|\cosmoset) \, .
    \label{eqn:total_GW_events_likelihood}
\end{equation}

The single GW event quasilikelihood is obtained through the marginalization over the source parameters. By defining $\vector{x}=\{\dl, \, z, \, \theta, \, \varphi \}$, we can write

\begin{equation}
    \likelihood(gw_i|\cosmoset) = \int d\vector{x} \, \likelihood(gw_i|\vector{x}, \cosmoset) \prior(\vector{x}|\cosmoset) \, .
\end{equation}

As in \cite{del2018stellar}, we assume that the integral over the sky position can be performed analytically. Hence we are left with

\begin{equation}
    \likelihood(gw_i|\cosmoset) = \int d\dl dz \,  \likelihood(gw_i|\dl, z, \cosmoset) \prior(\dl|z, \cosmoset) \prior(z|\cosmoset) \, .
\end{equation}

As shown in \cref{eqn:dl_z}, a given cosmological model fixes a $\dl$-$z$ relation. Therefore, among $\dl$, $z$ and $\cosmoset$, only two of them are independent quantities. We choose $\dl$ to be the dependent one, and the prior on the luminosity distance becomes, as in \cite{del2018stellar,del2012inference}, the following Dirac's delta

\begin{equation}
    \prior(\dl|z, \cosmoset) = \delta(\dl - \dl(z, \cosmoset)) \, ,
\end{equation}
where $\dl(z, \cosmoset)$ is computed through \cref{eqn:dl_z}. Under the Fisher information matrix formalism, the likelihood of a GW event is a Gaussian distribution in luminosity distance, whose mean value $\braket{\dl}$ and width $\sigma_{\dl}$ are given by the matched filtering technique. In addition to the instrument uncertainty of the luminosity distance, we also account for the systematic error due to weak-lensing, as modeled through $\sigma_{WL}$ (see eq.~(7.3) of~\cite{Tamanini:2016zlh}). Therefore, once we marginalize over the luminosity distance, we obtain

\begin{equation}
    \likelihood(gw_i|\dl(z, \cosmoset), z, \cosmoset) \propto \exp\Biggl[-\frac{1}{2} \frac{\bigl(\dl(z, \cosmoset) - \braket{\dl}\bigr)^2}{\sqrt{\sigma_{\dl}^2 + \sigma_{WL}^2}} \Biggr] \, .
\end{equation}

Finally, as in \cite{del2018stellar}, the prior over the GW event redshift is built in order to take into account for the peculiar velocities of galaxies in the catalog. Each galaxy $j$ is assigned with a hosting probability $w_j$ proportional to the distance in the ($\theta, \varphi$) plane between the host candidate and the relocated GW event. In particular, this quantity is computed by marginalization of \cref{eqn:error_distribution} over the luminosity distance \cite{laghi2021gravitational}

\begin{equation}
    w_{j} = \int \, d \dl \, p({\bf{\Theta}}|s) \, ,
\end{equation}
with ${\bf{\Theta}}=\{\dl, \mu_{N}, \varphi_{N}\}$, and by evaluating it at ($\mu_{N_{j}}, \phi_{N_{j}}$), which are the sky coordinates of the $j$-th galaxy within the error volume of a GW event. \par
The prior over the GW event redshift is therefore chosen as a discrete sum of $K$ Gaussians, to account for the galaxy redshift uncertainty, each weighted by its $w_j$ value:

\begin{equation}
    \prior(z|\cosmoset) \propto \sum_{j = 1}^{K} w_j \exp\biggl[ -\frac{1}{2} \biggl(\frac{z - z_j}{\sigma_{pv_{j}}} \biggr)^2 \biggr] \, .
    \label{eqn:redshift_prior}
\end{equation}

Here $j$ runs over the $K$ galaxies inside the error box, while $\sigma_{pv_{j}} = \sigma_{pv}(z_{j})$. The single GW quasilikelihood then reads

\begin{equation}
\begin{split}
    \likelihood(gw_i|\cosmoset) &\propto \int_{z_{\text{min}}}^{z_{\text{max}}} dz \, \exp{\Biggl[-\frac{1}{2} \frac{\bigl(\dl(z, \cosmoset) - \braket{\dl}\bigr)^2}{\sqrt{\sigma_{\dl}^2 + \sigma_{WL}^2}} \Biggr]} \times \\
    & \sum_{j = 1}^{N} w_j \exp{\biggl[ - \frac{1}{2} \biggl( \frac{z - z_j}{\sigma_{pv_{j}}} \biggr)^2 \biggr]} \, ,
\end{split}
    \label{eqn:single_GW_likelihood}
\end{equation}
where $z_{\text{min}}$ and $z_{\text{max}}$ are the lower and upper integration bounds and correspond to the minimum and maximum GW redshift obtained from the prior on $\cosmoset$ and inverting the $\dl$-$z$ relation in \cref{eqn:dl_z}.

\subsubsection{Prior choices and application}

In this work, we highlight the need of an independent estimate of the local universe expansion rate due to the current tension between the most recent estimates from EM surveys \cite{riess2019large,aghanim2020planck}. We therefore choose to infer a set of two different cosmological parameters which count the Hubble constant~\footnote{We define $h = \hubble/\SI{100}{\kilo\meter\per\second\per\mega\parsec}$, so that it is dimensionless and smaller than unity, in our case $h = 0.73$.} and the matter energy density parameter, $\cosmoset = \{h, \Om\}$. We choose conservative flat prior distributions for each quantity, in particular $h \in [0.6, 0.86]$ and  $\Om \in [0.04, 0.5]$. \par
The numerical implementation of the method described in \cref{bayes} is achieved through \texttt{COSMOLISA}, a public software package~\cite{cosmolisa} based on a nested sampling algorithm \cite{cpnest}. The primary output of nested sampling is the evidence $\evidence$, producing samples of the posterior distribution as a side product (see \cite{skilling2006nested} for the basics of the method). We explore the parameter space with \num{5000} live points and evolve, at each iteration, the lowest-likelihood one through a \emph{Markov Chain Monte Carlo} (MCMC) until $\evidence$ is computed at a given accuracy level.

\section{Results and discussion} \label{results}

\subsection{multiband approach} \label{multi_band}

We analysed the results of the parameter estimation for the three binary subpopulations (below-gap, across-gap, above-gap) and the four scenarios (mSFR-mZ, mSFR-sZ, sSFR-mZ, sSFR-sZ). As \cref{tab:detection} shows, however, not all the cases are suitable multiband candidates for the inference of cosmological parameters. For instance, across-gap binaries are extremely rare systems regardless of the scenario, while below-gap binaries reach interesting numbers only assuming \num{10} years of observation and the most optimistic models. As a matter of fact, above-gap binaries are the most promising ones, in particular under the sSFR-sZ scenario. Even if the corresponding intermediate models are still encouraging, in this work we focus on the optimistic one, as we need to impose other cuts on the catalog that further shrink the sample of useful systems. \par

\begin{table}
\centering
\begin{ruledtabular}
\begin{tabular}{ l c c c c c }
 & & mSFR-mZ & mSFR-sZ & sSFR-mZ & sSFR-sZ \\
\hline
\multirow{3}*{\num{4} years} & below-gap & \num{1.6} & \num{1.5} & \num{2.1} & \num{2.1} \\
& across-gap & \num{0.1} & \num{0.3} & \num{0.2} & \num{0.7}\\
& above-gap & \num{1.1} & \num{7.7} & \num{8.1} & \num{40.8} \\
\hline
\multirow{3}*{\num{10} years} & below-gap & \num{5.1} & \num{5.5} & \num{7.3} & \num{8.9} \\
& across-gap & \num{0.2} & \num{0.7} & \num{0.5} & \num{3.1}\\
& above-gap & \num{4.2} & \num{28.4} & \num{27.8} & \num{134.1} \\
\end{tabular}
\end{ruledtabular}
\caption{A showcase of the number of LISA's detection for a given population of binaries and model, assuming 4 and 10 years of mission lifetimes. These numbers represent one over the \num{30} realizations. Due to ET sensitivity curve, all binaries detected by LISA are also detected at ground.}
\label{tab:detection}
\end{table}

\begin{figure}
    \centering
    \includegraphics[width=0.5\textwidth]{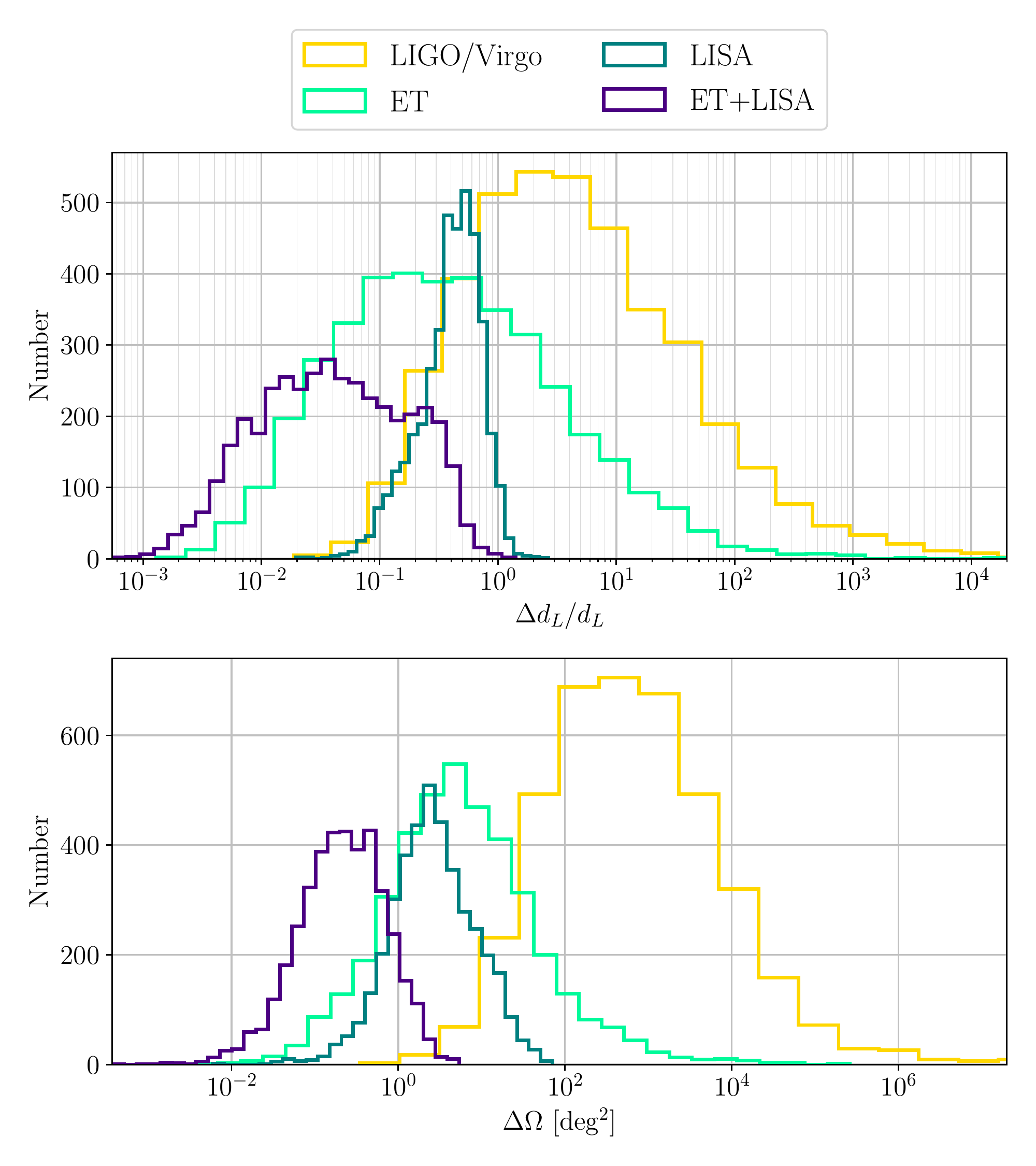}
    \caption{Uncertainty distributions for the luminosity distance (upper panel) and the sky localization (bottom panel) in LIGO/Virgo, ET, LISA and the network ET+LISA, colors as in legend. The panels display the results assuming \num{10} years LISA mission time and the above-gap, sSFR-sZ scenario.}
    \label{fig:fisher_errors}
\end{figure}

The first remarkable results are the high performances of the cooperation between space-born and third generation, ground-based interferometers. In \cref{fig:fisher_errors} we report the uncertainty distributions for $\dl$ and $\Delta\Omega$ for the above-gap and the sSFR-sZ model, assuming 10 years of LISA mission time. We consider LISA, ET and  LIGO-Virgo network at design sensitivity. Since the latter does not yield sufficient accuracy in parameter estimation to perform meaningful cosmological measurements, we focus on ET and LISA. Individually, both detectors hardly achieve the precision reached through their joint exploitation: for the sky location, the median of the ET+LISA distribution, in fact, decreases by an order of magnitude with respect to the single detector ones, with LISA being slightly better in the determination of this parameter. The same improvement concerns the luminosity distance, with ET being the best probe to measure this quantity, as expected.

In \cref{tab:medians} we show the medians of the uncertainty distributions for each parameter in the single detectors and in the network ET+LISA. Crucially for this work, ET+LISA yields an improvement factor of 100 in sky localization and of 10 in luminosity distance measurement precision. It should be noticed that the combination of the two also improves the estimate of all other parameters, most noticeably spin magnitudes.

\begin{table}[b]
\centering
\begin{ruledtabular}
\begin{tabular}{ l  c c  c  c }
 & \textbf{LIGO/Virgo} & \textbf{ET} & \textbf{LISA} & \textbf{ET+LISA} \\
\hline
$\Delta\Omega$ [\si{\square\degree}] & \num{5.6e2} & \num{5.1} & \num{2.5} & \num{2.2e-1} \\
$\Delta \dl / \dl$ & \num{3.6} & \num{3.4e-1} & \num{4.3e-1} & \num{3.9e-2} \\
$\Delta \mchirp / \mchirp$ & \num{3.6e-1} & \num{3.3e-3} & \num{4.1e-7} & \num{2.0e-7}\\
$\Delta \iota / \iota$ & \num{5.9} & \num{4.7e-1} & \num{9.3e-1} & \num{6.5e-2} \\
$\Delta \chi_1 / \chi_1$ & \num{3.9} & \num{2.8e-2} & \num{3.3e-1} & \num{8.6e-4} \\
$\Delta \chi_2 / \chi_2$ & \num{3.8} & \num{3.0e-2} & \num{4.6e-1} & \num{1.4e-3} \\
\end{tabular}
\end{ruledtabular}
\caption{Medians of the uncertainty distributions of each parameter for LIGO/Virgo, ET, LISA and the network ET+LISA, assuming \num{10} years LISA mission time. Here $\mchirp$ represents the \emph{chirp mass} of the system, defined as $\mathcal{M} = (M_1M_2)^{(3/5)}/(M_1+M_2)^{(1/5)}$  and its uncertainty is obtained through error propagation formulas.}
\label{tab:medians}
\end{table}

\subsection{Inference of cosmological parameters} \label{inference}

Among the many standard siren candidates, we impose a few cuts on the original binary catalogs. First, we request that the luminosity distance of the event is determined within $\num{10}\%$ of precision. Then we select only the events localized with $\Delta\Omega <  \SI{1}{\square\degree}\,$~\footnote{The likelihood computation becomes too expensive for GW events with more than $10^4$ hosts. Those events are generally poorly localized and their large 3D volumes result in a rather uninformative host redshift distribution.}. Moreover, due to the limited light cone extension, we request that the maximum redshift of the error box when varying the cosmology is smaller than \num{1}. Considering \num{4} (\num{10}) years of LISA observations, in 30 independent realizations of the Universe, from a total of \num{1223} (\num{4023}) binaries we are left with \num{222} (\num{510}) GW events, as reported in  \cref{tab:number_of_events_for_inference}.

\begin{table}
\centering
\begin{ruledtabular}
\begin{tabular}{l c c c}
Number of events: & Before cuts & After cuts & Per realization \\
\num{4} years & \num{1223} & \num{222} & $\sim 7$ \\
\num{10} years & \num{4023} & \num{510} & $\sim 17$ \\
\end{tabular}
\end{ruledtabular}
\caption{Number of events expected before and after the requirements imposed on the above-gap/sSFR-sZ binary catalog, assuming both LISA mission lifetimes. The rightmost column shows the average number of events per realization once we performed the selections.}
\label{tab:number_of_events_for_inference}
\end{table}

\subsubsection{Preliminary tests} \label{pre_test}

\begin{figure*}[t]
    \centering
    \includegraphics[width=0.4\textwidth]{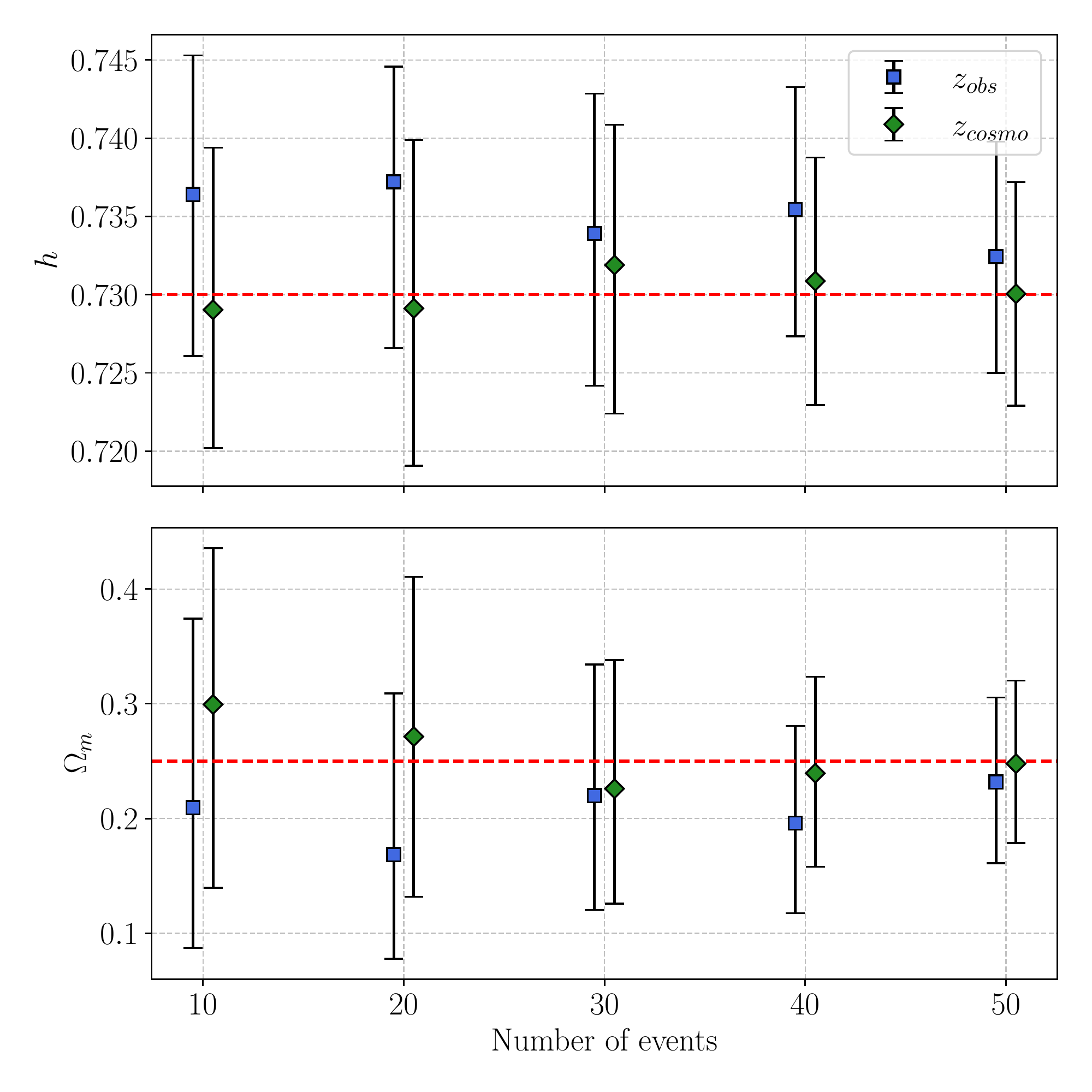}
    \includegraphics[width=0.4\textwidth, height=0.39\textwidth]{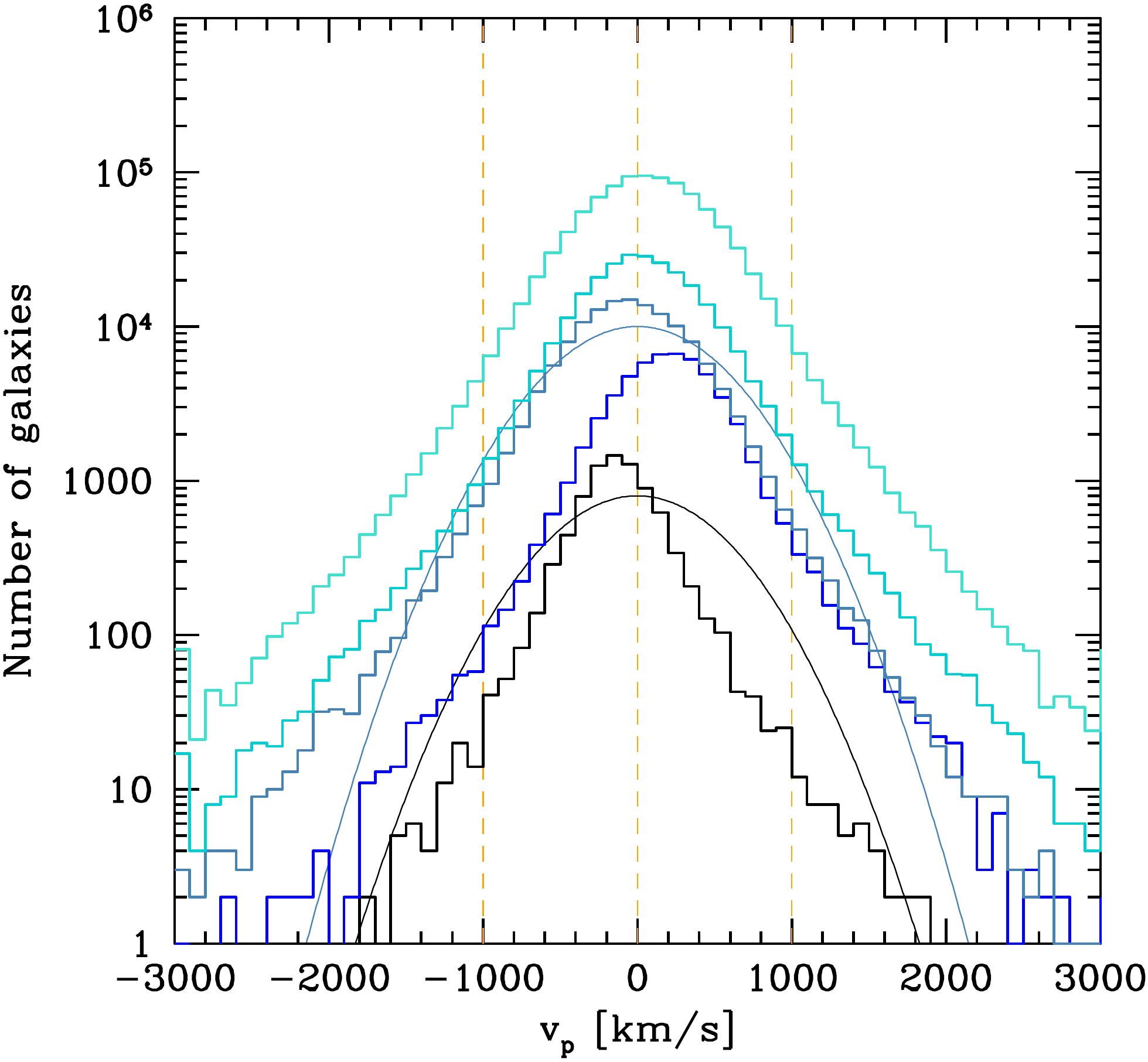}
    \caption{Left panel: cosmological parameter measurements as a function of the number of closest GW events used for the inference assuming a Universe with and without peculiar velocities (labeled $z_{obs}$ and $z_{cosmo}$, respectively). We investigate the results from a Universe with and without peculiar velocities, colors as in legend. The top plot represents the results for $h$, while the bottom plot displays the $\Om$ estimates (median values with error bars representing the $\num{1}\sigma$ confidence level). The red dashed line represents the true value of each parameter. Right panel: distribution of peculiar velocities in our light cone for several redshift bin. Histograms from bottom to top are shown for incremental \num{0.05} redshift bins, starting from [\num{0}, \num{0.05}]. To guide the eye, solid lines represent Gaussians with standard deviation $\sigma_{v_p}=500\,$km s$^{-1}$, as assumed in our study.}
    \label{fig:preliminary_test}
\end{figure*}

Since the analysis over a large number of events can be computationally costly, we start by exploring a small fraction of the total \num{10}-year catalog of events. We measure $h$ and $\Om$ by considering the closest GW events first, adding progressively farther events to the inference. \par
Even though the results are consistent with the true cosmology, we find estimates that are systematically biased toward large values of $h$ and small values of $\Om$, due to the correlation between the parameters. However, as the number of events increases, this effect is mitigated and eventually disappears, as the algorithm returns well centered Gaussian posterior distributions and the accuracy of the $h$ ($\Om$) measures evolves from \num{0.63}$\sigma$ (\num{0.28}$\sigma$) to \num{0.29}$\sigma$ (\num{0.24}$\sigma$). To understand whether this behavior is due to low redshift events only, we focus on these events in the following discussion. \par
The inference relies on the determination of the posterior distribution of each GW event redshift. If this step produces misleading information, the bias is propagated to the estimates of the cosmological parameters. To leading order, if the observed redshift is underestimated (overestimated), the $h$ and $\Om$ posteriors will be biased toward low (large) and high (low) values, respectively. The accuracy of our results suggests that the closer the GW event is, the more the observed redshift is overestimated. In light of these considerations, peculiar velocities may play a role in altering significantly the apparent redshift of close objects compared to the cosmological one. \par
As already mentioned in \cref{errbox}, the galaxy catalog comes with both the geometrical and the observed redshift of each object, thus allowing to assess the impact of the peculiar velocity on the analysis.
Through the same aforementioned procedure, we therefore infer the cosmological parameters in a Universe without peculiar velocities. The results are shown in the left panel of \cref{fig:preliminary_test}, where we directly compare the estimates from a static ($z_{cosmo}$) and dynamic ($z_{obs}$) Universe as a function of the number of closest GW events. As we can see, the bias completely vanishes already with \num{10} events, and as the number of mergers increases, the two analyses lead to the true cosmology. \par
Thus, peculiar velocities play a major role in the inference of cosmological parameters with low-redshift events. In fact, the right panel of \cref{fig:preliminary_test} shows that the $v_p$ distribution of our light cone galaxies is not symmetrically distributed around zero at low redshifts. The quantity $v_p$ is computed along the radial direction with positive values corresponding to a drift away from the observer. There is a clear preference for positive $v_p$ values for galaxies at $0.05<z<0.1$, which is where most of the closest GW events in our catalogs occur, thus explaining the bias. \par
It is interesting that our light cone features a large portion of galaxies that move preferentially away from the observer. One possible cause is the limited solid angle covered by our galaxy catalog. Although a full sky would mitigate this issue, its generation would be extremely computationally expensive. Here we just notice that the bias in the parameters was found considering the closest events of the full GW source catalog, featuring 30 realizations of the experiment. In a single realization there will be perhaps only one such close source. In fact, as we will see in \cref{real_observations}, this bias does not systematically appear in this case. 
Moreover, in a real experiment, one can in principle further mitigate any issue related to peculiar velocities by modelling the bulk motions as a function of redshift and update the inference model to keep into account for any local anisotropy.

\subsubsection{Individual realizations of the full experiment}\label{real_observations}

\begin{figure*}
    \centering
    \includegraphics[width=0.4\textwidth]{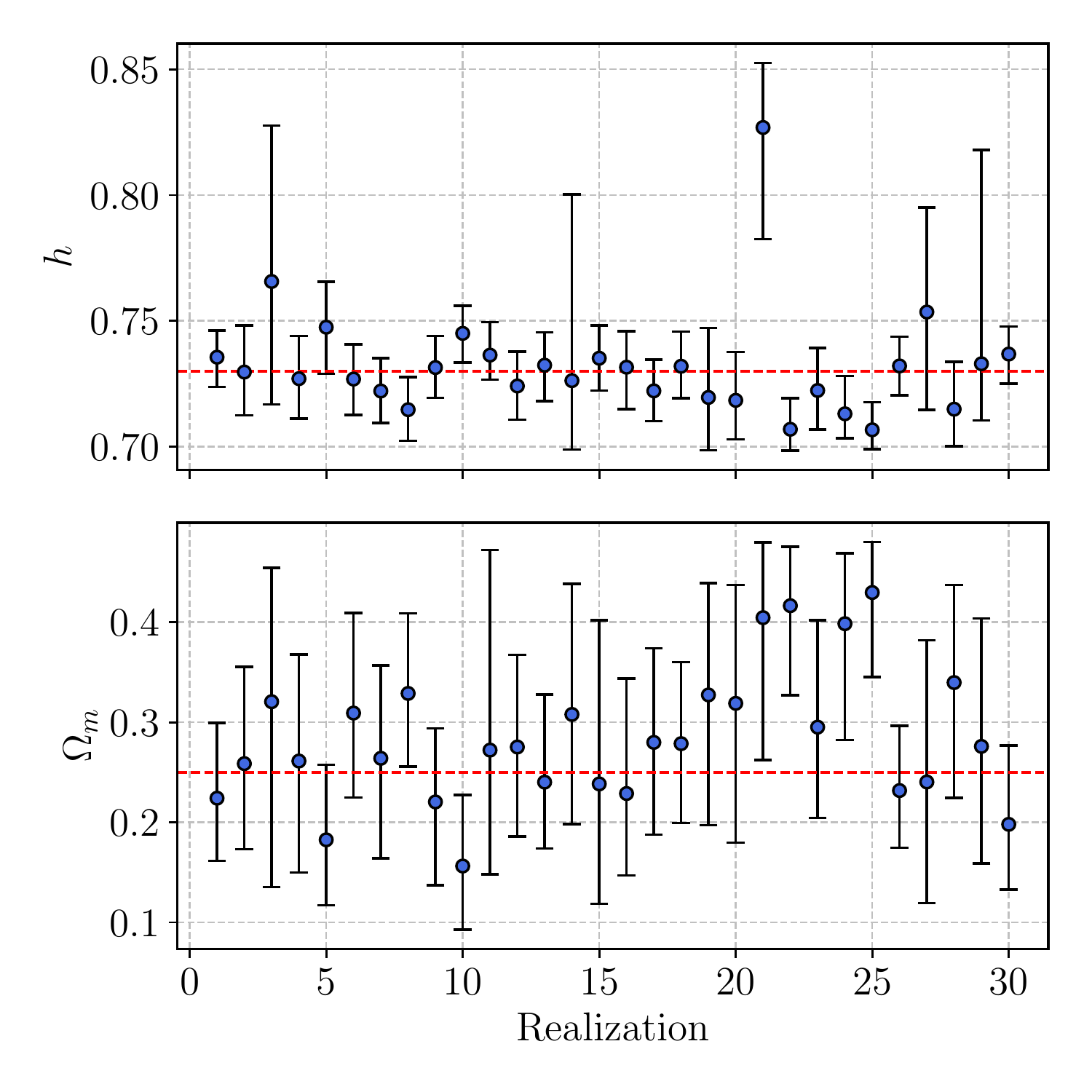}
    \includegraphics[width=0.4\textwidth]{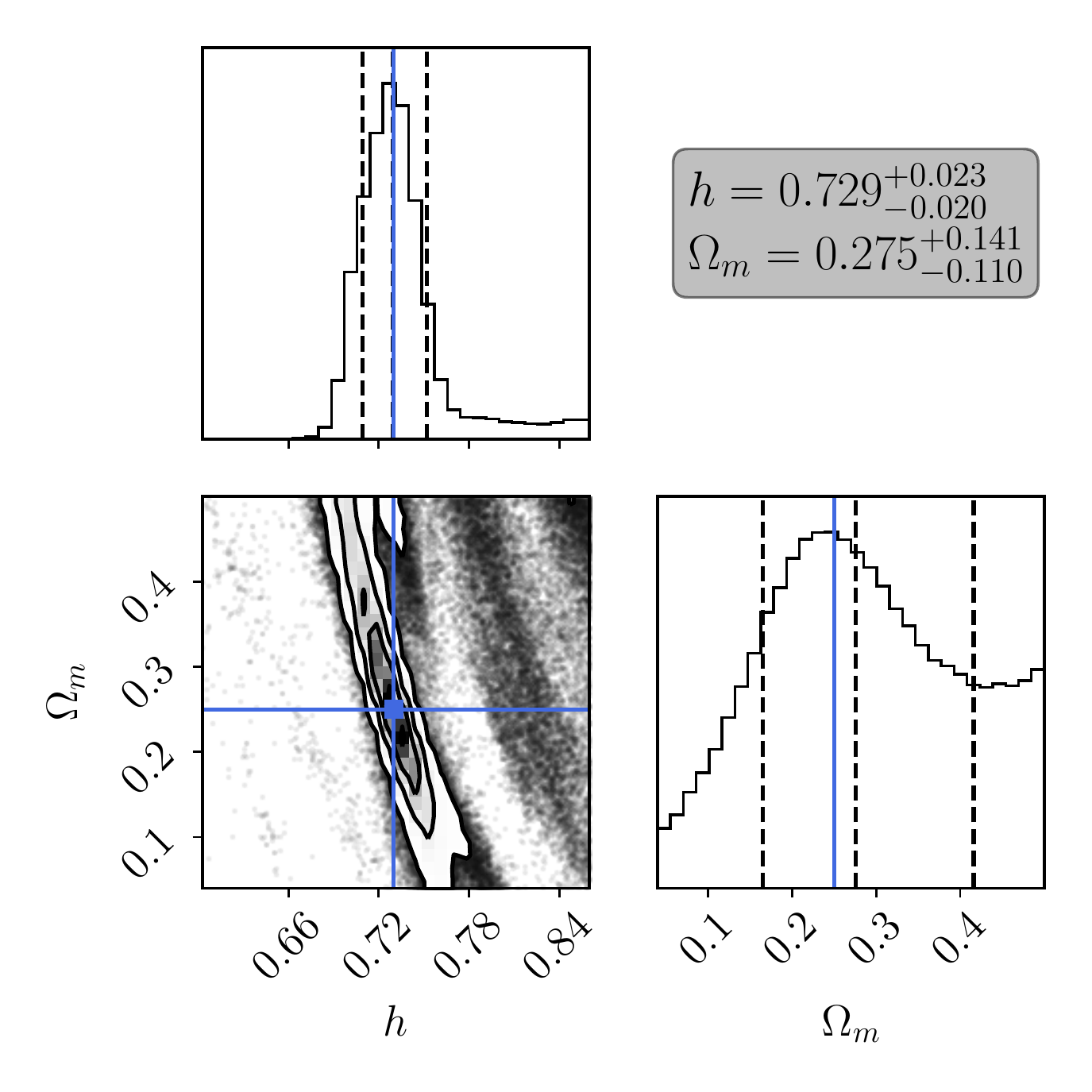} \\
    \includegraphics[width=0.4\textwidth]{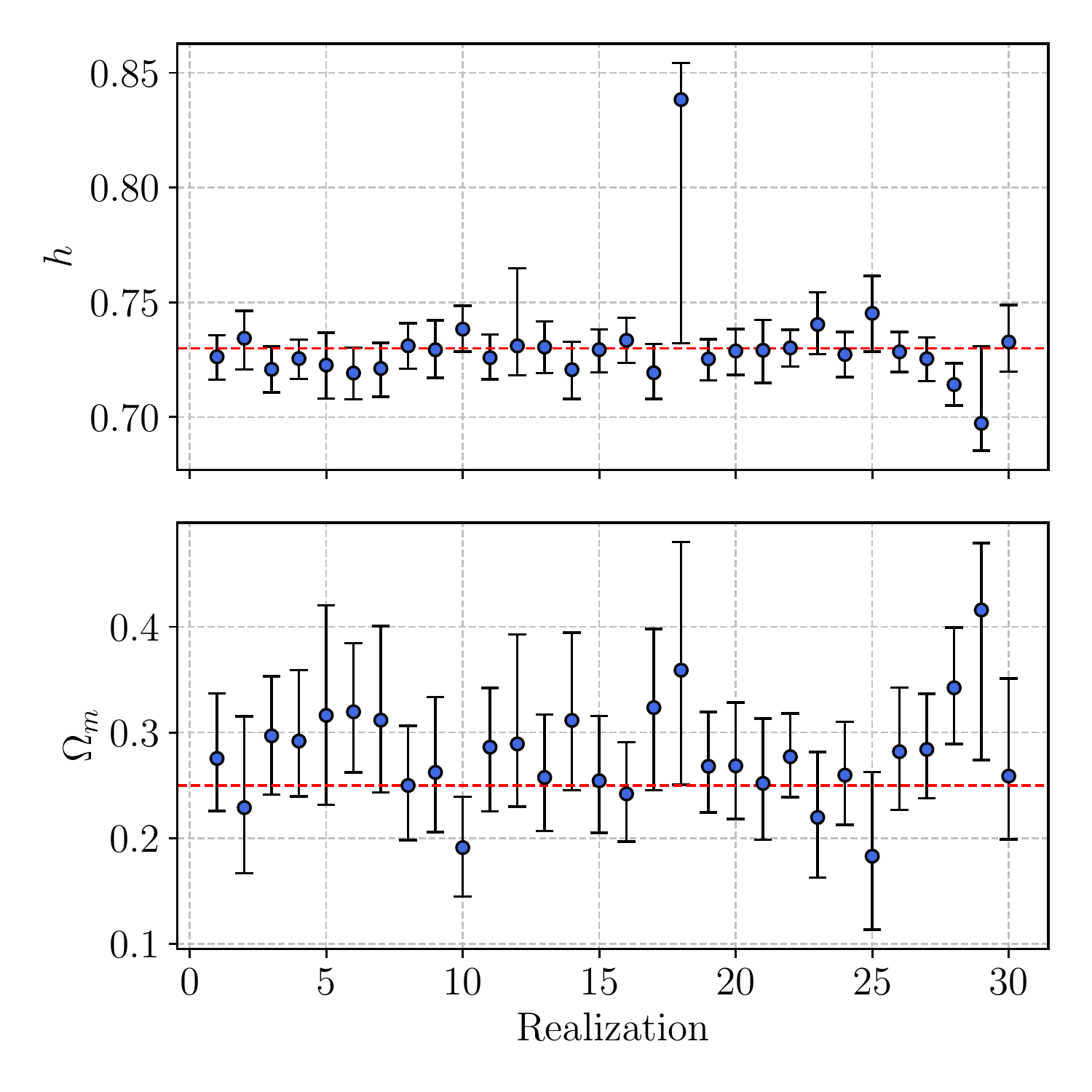}
    \includegraphics[width=0.4\textwidth]{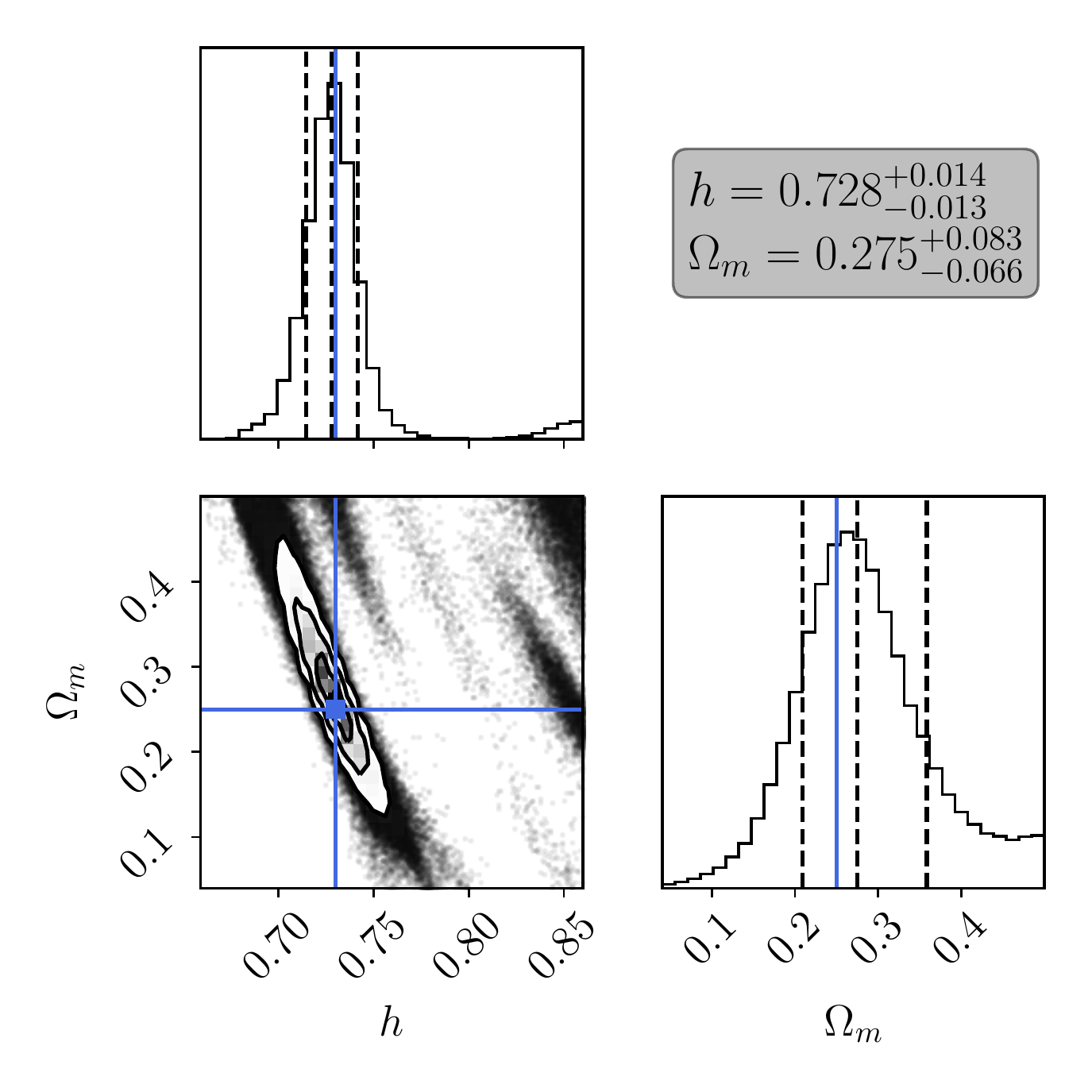}
    \caption{Inference results from the \num{30} independent realizations at the $68\%$ confidence level. The top row refers to \num{4} years of LISA observations, while the bottom row refers to \num{10} years. Plots in the left column, the blue dots denote the median of the posterior distributions, while the red dashed lines represent the true value of each parameter. The right column displays the joint posterior distributions averaged on the \num{30} independent realizations. Here the gray color scale distinguishes low probability regions (light) with high probability regions (dark); the black dashed lines mark, from left to right, the \num{16}, \num{50} and \num{84} percentile and finally the blue lines highlight the true cosmology. Plots in the right column have been made using \cite{corner}.}
    \label{fig:results}
\end{figure*}

\begin{table*}
\centering
\begin{ruledtabular}
\begin{tabular}{c c c c c c c c c} 
\multirow{2}*{LISA mission time} & \multirow{2}*{Realization} & \multirow{2}*{GW events} & \multicolumn{3}{c}{$h$} & \multicolumn{3}{c}{$\Om$} \\ 
 						         &	     	   &           &  $1\sigma$ &  $\%$  &  $A$  & $1\sigma$  &  $\%$ &  $A$     \\
\hline
\multirow{30}*{\num{4} years} & $1$ &  $8$  & $0.736^{+0.011}_{-0.012}$  & $1.5$  & $0.5\sigma$   &   $0.224^{+0.075}_{-0.063}$  & $30.74$  & $0.4\sigma$ \\
                              & $2$ &  $8$  & $0.730^{+0.019}_{-0.017}$  & $2.5$ & $< 0.1\sigma$  &  $0.259^{+0.097}_{-0.085}$  & $35.3$  & $0.1\sigma$  \\
                              & $3$ &  $8$  & $0.766^{+0.062}_{-0.049}$  & $7.3$  & $0.6\sigma$   &   $0.321^{+0.134}_{-0.185}$  & $49.7$  & $0.4\sigma$  \\
                              & $4$ &  $8$  & $0.727^{+0.017}_{-0.016}$  & $2.3$  & $0.2\sigma$    &  $0.261^{+0.106}_{-0.112}$  & $41.7$  & $0.1\sigma$  \\
                              & $5$ &  $8$  & $0.747^{+0.018}_{-0.018}$  & $2.4$  & $1.0\sigma$    &   $0.183^{+0.075}_{-0.065}$ & $38.5$  & $1.0\sigma$   \\
                              & $6$ &  $8$  & $0.727^{+0.014}_{-0.014}$  & $1.9$  & $0.2\sigma$   &   $0.309^{+0.100}_{-0.084}$  & $29.8$  & $0.6\sigma$  \\
                              & $7$ &  $8$  & $0.722^{+0.013}_{-0.013}$  & $1.8$  & $0.6\sigma$   &   $0.264^{+0.093}_{-0.100}$  & $36.5$  & $0.1\sigma$  \\
                              & $8$ &  $8$  & $0.715^{+0.013}_{-0.012}$  & $1.8$  & $1.2\sigma$   &   $0.329^{+0.080}_{-0.073}$  & $23.3$  & $1.0\sigma$  \\
                              & $9$ &  $8$  & $0.731^{+0.013}_{-0.012}$  & $1.7$  & $0.1\sigma$   &   $0.220^{+0.073}_{-0.083}$  & $35.5$  & $0.4\sigma$  \\
                              & $10$ & $8$  & $0.745^{+0.011}_{-0.012}$  & $1.5$  & $1.3\sigma$   &   $0.156^{+0.071}_{-0.063}$ & $42.9$  & $1.4\sigma$   \\
                              & $11$ &  $8$  & $0.736^{+0.013}_{-0.010}$  & $1.6$ & $0.6\sigma$   &  $0.272^{+0.200}_{-0.124}$  & $59.57$  & $0.1\sigma$  \\
                              & $12$ &  $8$  & $0.724^{+0.014}_{-0.013}$  & $1.9$ & $0.4\sigma$   &   $0.275^{+0.092}_{-0.089}$  & $32.9$  & $0.3\sigma$  \\
                              & $13$ &  $7$  & $0.732^{+0.013}_{-0.014}$  & $1.9$ & $0.2\sigma$    &   $0.240^{+0.087}_{-0.066}$  & $32.0$  & $0.1\sigma$  \\
                              & $14$ &  $7$  & $0.726^{+0.074}_{-0.027}$  & $7.0$ & $0.1\sigma$    &   $0.308^{+0.131}_{-0.109}$ & $39.0$  & $0.5\sigma$   \\
                              & $15$ &  $7$  & $0.735^{+0.013}_{-0.013}$  & $1.8$  & $0.4\sigma$   &   $0.238^{+0.164}_{-0.120}$  & $59.5$  & $0.1\sigma$  \\
                              & $16$ &  $7$  & $0.732^{+0.014}_{-0.017}$  & $2.1$  & $0.1\sigma$   &   $0.229^{+0.115}_{-0.082}$  & $43.0$  & $0.2\sigma$  \\
                              & $17$ &  $7$  & $0.722^{+0.012}_{-0.012}$  & $1.7$  & $0.6\sigma$   &   $0.280^{+0.094}_{-0.092}$  & $33.3$  & $0.3\sigma$  \\
                              & $18$ &  $7$  & $0.732^{+0.014}_{-0.013}$  & $1.8$  & $0.1\sigma$   &   $0.279^{+0.081}_{-0.079}$  & $29.0$  & $0.4\sigma$  \\
                              & $19$ & $7$   & $0.720^{+0.028}_{-0.021}$  & $3.4$  & $0.4\sigma$   &   $0.327^{+0.112}_{-0.130}$ & $36.9$  & $0.6\sigma$   \\
                              & $20$ &  $7$  & $0.718^{+0.019}_{-0.015}$  & $2.4$  & $0.7\sigma$   &  $0.319^{+0.118}_{-0.139}$  & $40.4$  & $0.5\sigma$  \\
                              & $21$ &  $7$  & $0.827^{+0.026}_{-0.044}$  & $4.2$  & $2.8\sigma$   &   $0.405^{+0.075}_{-0.142}$  & $26.9$  & $1.4\sigma$  \\
                              & $22$ &  $7$  & $0.707^{+0.012}_{-0.008}$  & $1.5$  & $2.2\sigma$   &   $0.416^{+0.059}_{-0.090}$  & $17.8$  & $2.2\sigma$  \\
                              & $23$ &  $7$  & $0.722^{+0.017}_{-0.016}$  & $2.2$  & $0.5\sigma$   &   $0.295^{+0.107}_{-0.091}$ & $33.5$  & $0.5\sigma$   \\
                              & $24$ &  $7$  & $0.713^{+0.015}_{-0.010}$  & $1.7$  & $1.4\sigma$   &   $0.398^{+0.0.70}_{-0.116}$  & $23.4$  & $1.6\sigma$  \\
                              & $25$ &  $7$  & $0.707^{+0.011}_{-0.008}$  & $1.3$  & $2.5\sigma$   &   $0.430^{+0.051}_{-0.084}$  & $15.7$  & $2.7\sigma$  \\
                              & $26$ &  $7$  & $0.732^{+0.012}_{-0.012}$  & $1.6$  & $0.2\sigma$   &   $0.232^{+0.065}_{-0.057}$  & $26.2$  & $0.3\sigma$  \\
                              & $27$ &  $7$  & $0.754^{+0.042}_{-0.039}$  & $5.3$  & $0.6\sigma$   &   $0.240^{+0.142}_{-0.121}$  & $54.6$  & $0.1\sigma$  \\
                              & $28$ & $7$  & $0.715^{+0.019}_{-0.015}$  & $2.4$.  & $0.9\sigma$   &   $0.340^{+0.097}_{-0.115}$ & $31.3$  & $0.8\sigma$   \\
                              & $29$ &  $7$  & $0.733^{+0.085}_{-0.022}$  & $7.3$  & $0.1\sigma$   &   $0.276^{+0.128}_{-0.117}$  & $44.4$  & $0.2\sigma$  \\
                              & $30$ & $7$  & $0.737^{+0.011}_{-0.012}$  & $1.5$   & $0.6\sigma$   &   $0.198^{+0.079}_{-0.065}$ & $36.4$  & $0.7\sigma$   \\
    \hline
\multirow{30}*{\num{10} years} & $1$ &  $17$  & $0.726^{+0.009}_{-0.010}$  & $1.3$ & $0.4\sigma$   &   $0.275^{+0.062}_{-0.050}$  & $22.2$  & $0.5\sigma$   \\
                              & $2$ &  $17$  & $0.734^{+0.012}_{-0.014}$  & $1.7$  & $0.3\sigma$   &  $0.229^{+0.086}_{-0.062}$  & $32.4$  & $0.3\sigma$  \\
                              & $3$ &  $17$  & $0.721^{+0.010}_{-0.010}$  & $1.4$  & $0.9\sigma$   &   $0.297^{+0.056}_{-0.055}$  & $18.9$  & $0.8\sigma$  \\
                              & $4$ &  $17$  & $0.725^{+0.008}_{-0.009}$  & $1.2$  & $0.5\sigma$   &   $0.292^{+0.067}_{-0.052}$  & $20.4$  & $0.7\sigma$  \\
                              & $5$ &  $17$  & $0.723^{+0.014}_{-0.015}$  & $2.0$   & $0.5\sigma$  &   $0.316^{+0.104}_{-0.085}$ & $28.9$   & $0.7\sigma$   \\
                              & $6$ &  $17$  & $0.719^{+0.011}_{-0.011}$  & $1.6$  & $1.0\sigma$   &   $0.320^{+0.065}_{-0.057}$  & $19.1$  & $1.1\sigma$  \\
                              & $7$ &  $17$  & $0.721^{+0.011}_{-0.012}$  & $1.6$  & $0.8\sigma$   &   $0.312^{+0.089}_{-0.068}$  & $25.3$  & $0.8\sigma$  \\
                              & $8$ &  $17$  & $0.731^{+0.010}_{-0.010}$  & $1.7$  & $0.1\sigma$   &   $0.250^{+0.056}_{-0.052}$  & $21.6$  & $< 0.1\sigma$  \\
                              & $9$ &  $17$  & $0.720^{+0.013}_{-0.012}$  & $1.7$ & $0.1\sigma$   &   $0.262^{+0.071}_{-0.056}$  & $24.3$  & $0.2\sigma$  \\
                              & $10$ & $17$  & $0.738^{+0.010}_{-0.010}$  & $1.4$  & $0.8\sigma$   &   $0.191^{+0.048}_{-0.046}$ & $24.6$  & $1.3\sigma$   \\
                              & $11$ &  $17$  & $0.726^{+0.010}_{-0.009}$  & $1.4$ & $0.4\sigma$   &  $0.286^{+0.056}_{-0.061}$  & $20.4$  & $0.6\sigma$  \\
                              & $12$ &  $17$  & $0.731^{+0.034}_{-0.013}$  & $3.2$ & $< 0.1\sigma$   &   $0.289^{+0.103}_{-0.059}$  & $28.1$  & $0.5\sigma$  \\
                              & $13$ &  $17$  & $0.730^{+0.011}_{-0.011}$  & $1.5$ & $< 0.1\sigma$  &   $0.257^{+0.060}_{-0.051}$  & $21.4$  & $0.1\sigma$  \\
                              & $14$ &  $17$  & $0.721^{+0.012}_{-0.013}$  & $1.7$ & $0.8\sigma$.  &   $0.311^{+0.083}_{-0.066}$ & $23.9$  & $0.8\sigma$   \\
                              & $15$ &  $17$  & $0.729^{+0.009}_{-0.010}$  & $1.3$  & $0.1\sigma$   &   $0.254^{+0.061}_{-0.049}$  & $21.7$  & $0.1\sigma$  \\
                              & $16$ &  $17$  & $0.733^{+0.010}_{-0.010}$  & $1.3$  & $0.3\sigma$   &   $0.242^{+0.049}_{-0.045}$  & $19.4$  & $0.2\sigma$  \\
                              & $17$ &  $17$  & $0.719^{+0.012}_{-0.011}$  & $1.7$  & $0.9\sigma$   &   $0.324^{+0.074}_{-0.078}$  & $23.6$  & $1.0\sigma$  \\
                              & $18$ &  $17$  & $0.838^{+0.016}_{-0.106}$  & $7.3$  & $1.8\sigma$   &   $0.359^{+0.121}_{-0.109}$  & $32.0$  & $0.9\sigma$  \\
                              & $19$ & $17$  & $0.725^{+0.009}_{-0.009}$   & $1.2$  & $0.5\sigma$   &   $0.268^{+0.051}_{-0.044}$ & $17.7$  & $0.4\sigma$   \\
                              & $20$ &  $17$  & $0.729^{+0.010}_{-0.010}$  & $1.4$  & $0.1\sigma$   &  $0.268^{+0.060}_{-0.050}$  & $20.5$  & $0.3\sigma$  \\
                              & $21$ &  $17$  & $0.729^{+0.013}_{-0.014}$  & $1.9$  & $0.1\sigma$   &   $0.252^{+0.061}_{-0.053}$  & $22.8$  & $< 0.1\sigma$  \\
                              & $22$ &  $17$  & $0.730^{+0.008}_{-0.008}$  & $1.1$  & $<0.1\sigma$  &   $0.277^{+0.041}_{-0.038}$  & $14.3$  & $0.7\sigma$  \\
                              & $23$ &  $17$  & $0.740^{+0.014}_{-0.013}$  & $1.8$  & $0.8\sigma$   &   $0.220^{+0.062}_{-0.057}$ & $27.0$  & $0.5\sigma$   \\
                              & $24$ &  $17$  & $0.727^{+0.010}_{-0.010}$  & $1.4$  & $0.3\sigma$   &   $0.260^{+0.051}_{-0.047}$  & $18.8$  & $0.2\sigma$  \\
                              & $25$ &  $17$  & $0.745^{+0.016}_{-0.017}$  & $2.2$  & $0.9\sigma$   &   $0.183^{+0.079}_{-0.069}$  & $40.7$  & $0.9\sigma$  \\
                              & $26$ &  $17$  & $0.728^{+0.009}_{-0.009}$  & $1.2$  & $0.2\sigma$   &   $0.282^{+0.061}_{-0.055}$  & $20.5$  & $0.6\sigma$  \\
                              & $27$ &  $17$  & $0.725^{+0.009}_{-0.010}$  & $1.3$  & $0.5\sigma$   &   $0.284^{+0.053}_{-0.046}$  & $17.4$  & $0.7\sigma$  \\
                              & $28$ & $17$  & $0.714^{+0.009}_{-0.009}$  & $1.3$   & $1.7\sigma$   &   $0.342^{+0.057}_{-0.053}$ & $16.1$  & $1.7\sigma$   \\
                              & $29$ &  $17$  & $0.697^{+0.034}_{-0.012}$  & $3.3$  & $1.4\sigma$   &   $0.416^{+0.063}_{-0.142}$  & $24.7$  & $1.6\sigma$  \\
                              & $30$ & $17$  & $0.733^{+0.016}_{-0.013}$  & $2.0$   & $0.2\sigma$  &   $0.259^{+0.092}_{-0.060}$ & $29.4$  & $0.1\sigma$   \\
\end{tabular}
\end{ruledtabular}
\caption{The $h$ and $\Om$ estimates for all the inference runs. The columns report: the estimates at the $68\%$ confidence level ($1\sigma$), the precision of the measure ($\%$), and the estimate accuracy as a fraction of $\sigma$ ($A$).}
\label{tab:results}
\end{table*}

The nominal LISA mission lifetime is set to be \num{4} years with a possible operation extension up to \num{10} years \cite{amaro2017laser}. To produce realistic samples of GW events for a given mission time, we divide the original catalog in independent subsets producing:

\begin{itemize}
    \item \num{30} realizations of either 7 or 8 events each for \num{4} years of observations; 
    \item \num{30} realizations of 17 events each for \num{10} years of observations. 
\end{itemize}

The sizes of both samples are easily manageable from the computational point of view. We therefore consider all individual realizations and then average the posterior distributions from the \num{30} subsets to characterize the accuracy and the precision of this method. \par
The results for the inference of $h$ and $\Om$ from all the runs are presented in \cref{fig:results,tab:results}. We observe that the bias due to peculiar velocities does not pose a significant issue in individual realizations of the experiment because of the very small number of low redshift events. In the vast majority of the cases, the inference yields measurements of $h$ that are both precise and accurate. By looking at \cref{tab:results} we can in fact appreciate that $h$ is generally measured to better than $2\%$ ($68\%$ credible region) and often around $1\%$ in the 10-year case, and the true value is generally well within the $68\%$ credible region implying no significant biases. $\Omega_m$ is far less constrained, generally at a 20-30\% level. \par
There are, however, a couple of realizations that yield severely inconsistent results. These outliers can be identified in realization 21 in the 4-year case and realizations 18 in the 10-year case (i.e.~in \num{1} of our \num{30} realizations). In these realizations, both $h$ and $\Omega_m$ rail against the upper end of the prior range, a behavior that is not explained by the expected bias due to peculiar velocities. The effect of this can also be seen in the averaged posteriors shown in the right panels of \cref{fig:results}. In the 4-year case the averaged marginalized posteriors of $h$ and $\Omega_m$ display long tails extending to the upper bound of the priors, whereas in the 10-year case this effect is largely suppressed. However, in this case a mild secondary peak appears at the boundary of the prior and leads to a slightly asymmetric distribution. Even if the averaged estimates that we obtain are largely consistent with the true cosmology, we want to focus our attention on the physical meanings (if any) of the small deviations from the expected results.

\subsubsection{Origin of the problematic realizations} \label{origin_prob_real}

\begin{figure}
    \centering
    \includegraphics[width=0.5\textwidth]{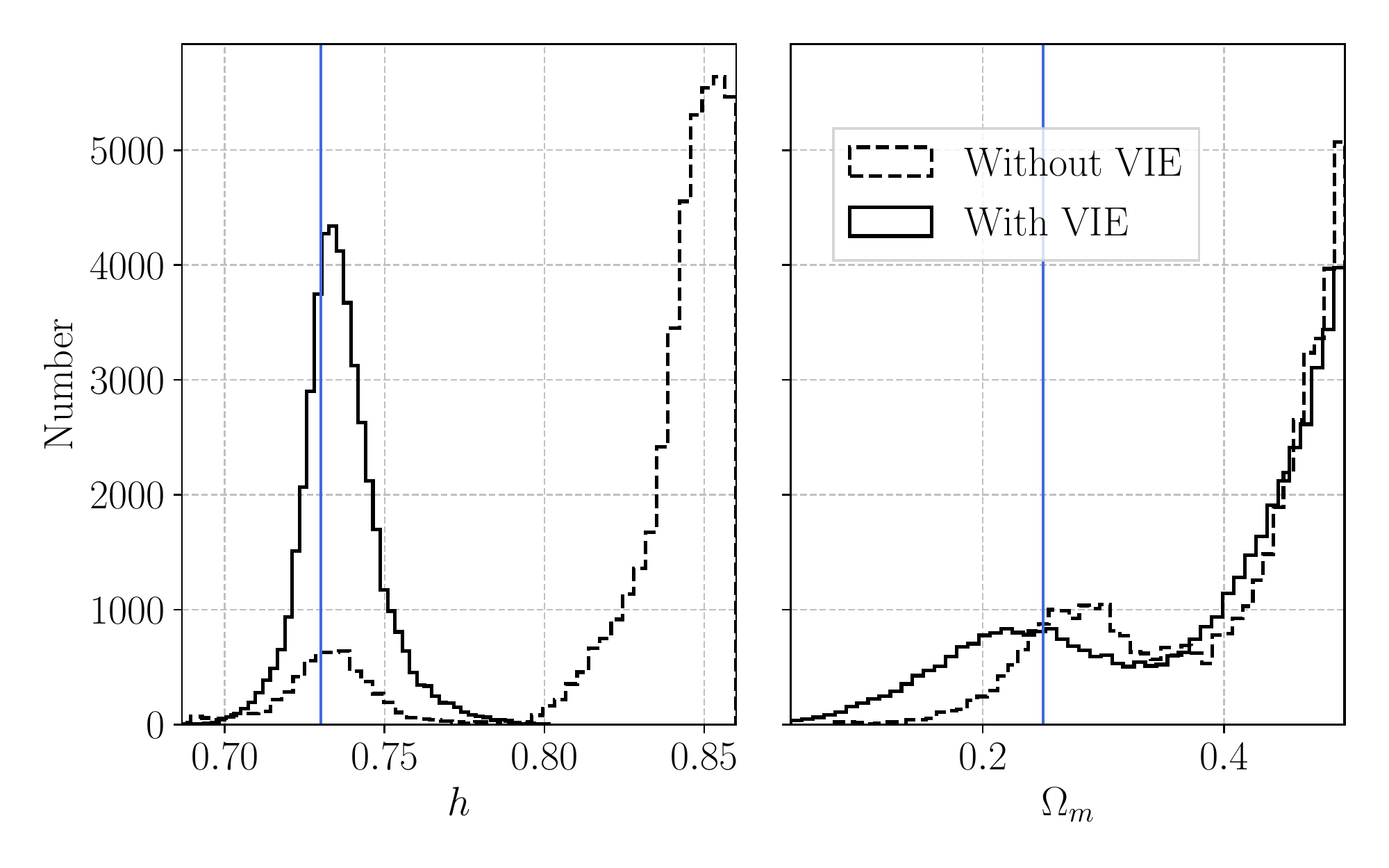}
    \caption{The impact of VIE on the $h$ and $\Om$ posterior distributions of realization \num{18} (\num{10} years). The blue solid lines mark the true cosmology.}
    \label{fig:before_after_event_174}
\end{figure}

\begin{figure}
    \centering
    \includegraphics[width=0.5\textwidth]{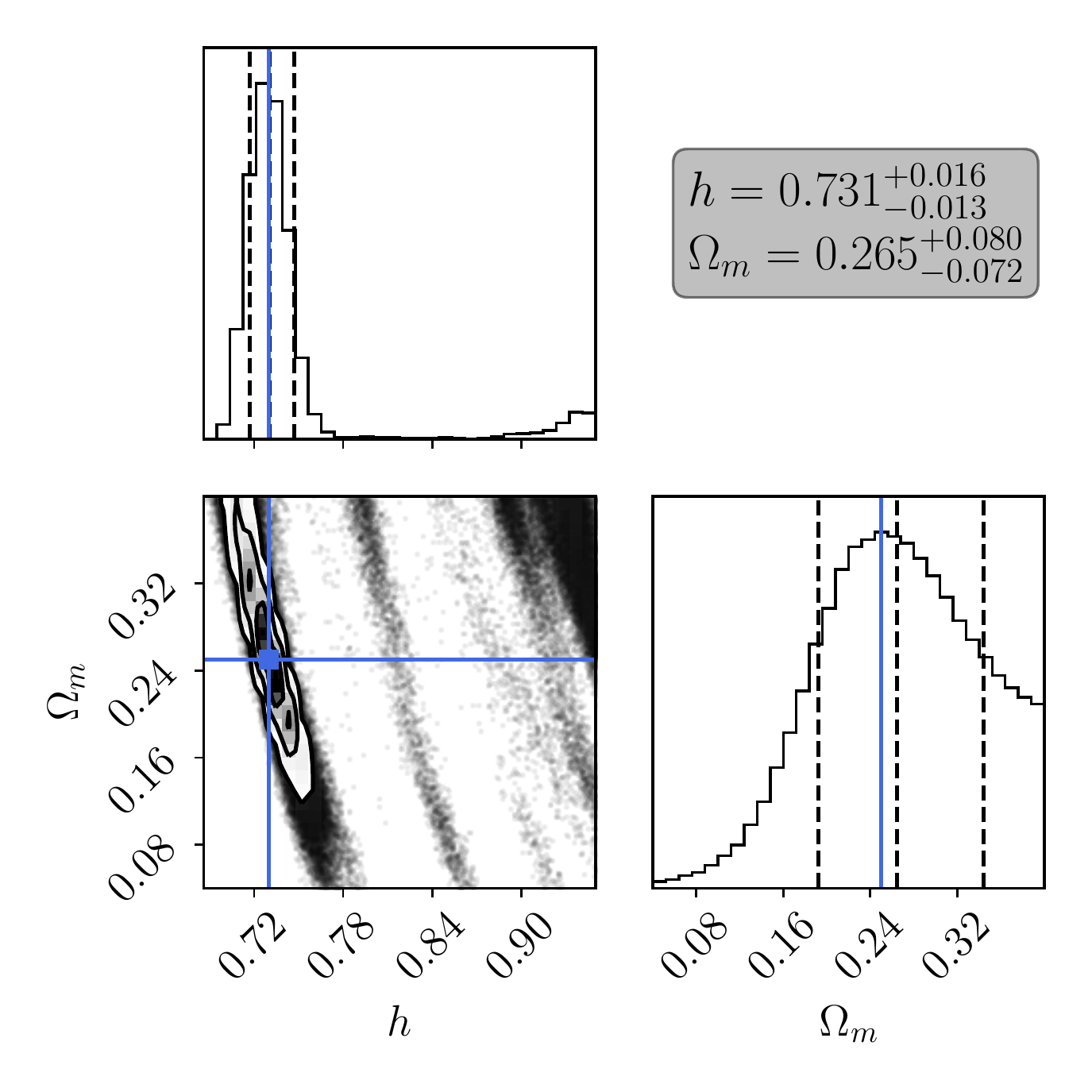}
    \caption{Averaged posterior distribution over \num{30} different realizations assuming different prior ranges on $h$ and $\Om$, as discussed in \cref{origin_prob_real}.}
    \label{fig:new_prior_ranges}
\end{figure}

We extensively examined the bad \num{10}-year realization and did not find any specific pathology in the error box construction, nor in the inference procedure.
The culprit of the bias appears to lie in the relation between $\hubble$, $\Om$ and redshift that enters the single-event likelihood \cref{eqn:single_GW_likelihood}, which is proportional to the number of galaxies at a given redshift. Therefore, for a comoving-volume-uniform galaxy distribution, the weight is roughly proportional to the comoving volume shell, naturally favoring values of redshift approaching $z_{\rm max}$. If the information enclosed in the clustering is not strong enough, this ``high $z$'' solution competes with the correct one and can eventually dominate. This issue is further exacerbated by the $\hubble - \Om$ degeneracy. As seen from the 2D posteriors shown in \cref{fig:results}, the two cosmological parameters are partially degenerate, showing a clear anticorrelation. Therefore, a given $(\dl, z)$ pair can be produced by a continuum of $(\hubble, \Om)$ following this degeneracy. For a rectangular uniform prior in the cosmological parameters, while galaxies in the middle of the error box are consistent with a degenerate set of cosmologies, those at the boundaries ($z_{\rm min}$ and $z_{\rm max}$) are only consistent with the corners of the prior. Since many more galaxies accumulate toward $z_{\rm max}$, this creates an artificial spike at the top right corner of the parameter space. So in absence of events with zero support around $z_{\rm max}$, this second mode of the posterior cannot be suppressed and would eventually dominate, essentially by design of the likelihood function. \par
The above interpretation is corroborated by two tests that we now discuss. The first test we performed was to artificially add to the bad \num{10}-year realization a very informative event (VIE), i.e.~an event with no (or little) support at redshifts outside those allowed by the true cosmology, in particular with zero support around $z_{\rm max}$. The result of this procedure is shown in \cref{fig:before_after_event_174}. It can be seen that the addition of a VIE kills the high $z$ solution and allows the recovery of the correct cosmology, albeit with large uncertainties on $\Om$. \par
A second test consisted in changing the prior range of the analysis. If our interpretation is correct, the second mode of the solution appearing in \cref{fig:results} should follow the boundary of the prior. \Cref{fig:new_prior_ranges} shows the average posterior over 30 realizations (different from those used to produce \cref{fig:results}) of the experiment with a modified prior range $h\in[0.6,0.95]$ and $\Om\in[0.04, 0.4]$. Clearly, the secondary mode follows the boundary of the prior, while the correct solution is consistently recovered regardless of the prior range.

\subsubsection{Mitigation techniques and future investigations}

\begin{figure*}
    \centering
    \includegraphics[width=1.\textwidth]{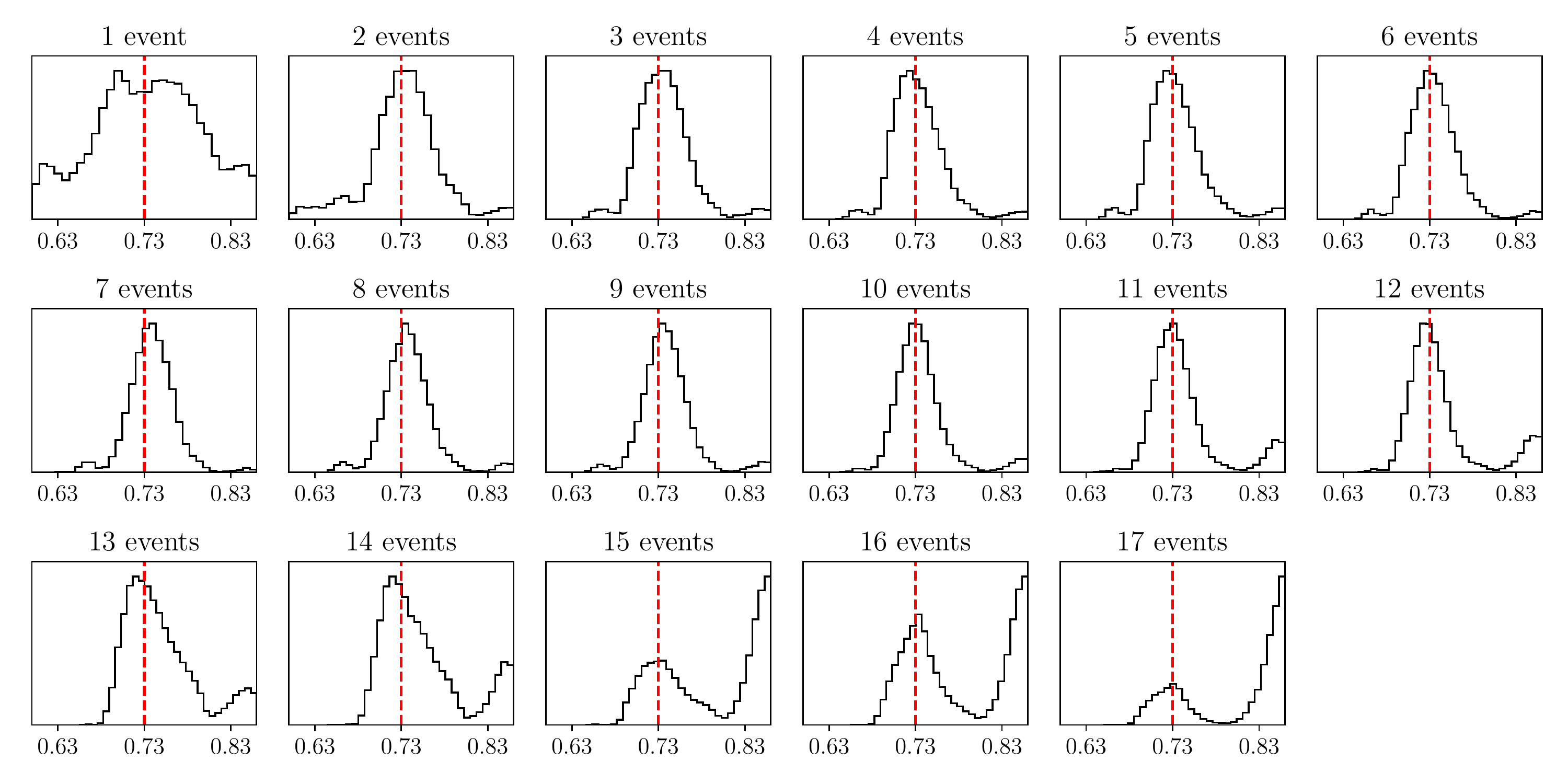}
    \caption{Evolution of the posterior distribution of the Hubble constant from the realization \num{18} (\num{10}-year) as the number of events increases by one each time, as the panel titles show. Here the red dotted line represents the true value, i.e.~$h=0.73$. In this particular case, the spurious solution overcomes the right cosmology when the last few events are added to the inference.}
    \label{fig:h_posterior_evolution}
\end{figure*}

Although we are focusing on the ``bad outcomes'' of our analysis, it is worth keeping in mind that they involve only a minority of realizations of the experiment. It is nevertheless important to be able to treat these potential issues should they manifest in a future real analysis on actual data. \par
Even without applying any change to the inference procedure, the tests presented above provide practical ways to reject spurious solutions and correctly constrain cosmological parameters. Bad realizations generally display bimodal posteriors. Repetition of the analysis with a varying prior range should return a ``steady mode'' around the correct solution and a dominant spurious peak following the boundary. Moreover, events can be analyzed individually and cosmological inference can be done progressively by adding them one by one. In order to exclude peculiar velocities' side effects, we perform this test by considering a steady Universe. As we show in \cref{fig:h_posterior_evolution} for realization \num{18}, we found that in our two bad realizations, when this procedure is employed, the joint posterior initially builds up around the correct solution due to the clustering information. This mode, however, is eventually superseded by the spurious one in the long run, if there are no events without host galaxy support around $z_{\rm max}$. These two checks (varying prior and incremental analysis) could allow us to reject the spurious solution and identify the correct cosmological parameters. \par
It would be obviously desirable to develop an analysis that prevents side effects like the ones identified here. For example one can reweight or change the shape of the prior, so that an uninformative experiment returns a flat posterior on the parameters, regardless of the fact that the assumption of a cosmological model intrinsically  brings some information on the correlation structure of the parameters. 
Another aspect that should be accounted for concerns the approach adopted to place each GW event in the Millennium Universe, that we discussed in \cref{errbox}. Within our work, the binary population and the galaxy catalog are independent entities, i.e.~binaries merge at points in space where there may not be any galaxy at all: to address the problem, we randomly selected the ``true host'' within the redshift interval associated to the true cosmology consistent with the $\dl$ uncertainty of the GW measurement. This ensures that an actual galaxy can be referred as the host of the GW event. The random selection should ensure that galaxies in denser regions are more likely to be drawn rather than others. However, the high performance of the multiband approach squeezes the pool of true host candidates to a narrow redshift interval, which often lacks of relevant clustering properties. Therefore, when we extend the redshift interval to take into account for the prior ranges on the cosmological parameters, we are likely to bring into the error box much denser regions that were artificially excluded from the host selection. In practice, one should start from a given light cone and place GW events randomly within it, to ensure that the distribution of events traces the 3D clustering of galaxies. We plan to explore different experiment designs and likelihood forms in future work. \par
Besides these adjustments, we remark once again that the analysis performed here allows a robust determination of $\hubble$ within \num{1}-\num{2}$\%$ in the vast majority of the cases. Moreover, even when the inference is biased, the spurious solution can be identified and the correct cosmology recovered. 

\section{Conclusions} \label{end}

In this paper we explored the possibility of exploiting multiband GW astronomy to use SBHBs above the pair-instability mass gap as effective dark standard sirens. \par
Massive SBHBs forming from progenitor stars above the pair-instability mass gap are in fact anticipated to be loud multiband sources, detectable both by LISA and 3G detectors, for which we considered ET as an example. By combining observations in the two bands, the source 3D sky localization can be pinned down to an accuracy which is far better (by three orders of magnitude, on average) than the individual probes alone. This allows an efficient probabilistic identification of the host among all galaxies within the error volume, enabling statistical inference on the cosmological parameters. \par
We exploited this idea to constrain the Hubble constant $\hubble$ and matter density fraction $\Om$ under the assumption of a flat, $\lcdm$ Universe. We performed the parameter estimation in each detector under the Fisher information matrix formalism, and then combined the relative uncertainties in order to produce accurate estimates of the source parameters, specifically the 3D localization error volume in the sky. We then relied on the Millennium simulation to simulate the galaxy distribution across \num{1}/\num{8}th of the sky up to $z=1$. We placed the 3D volumes in this synthetic sky, thus creating for each of them a sample of host candidates consistent with the clustering properties of the $\lcdm$ Universe. Finally, we used these data to perform Bayesian inference on $\hubble$ and $\Om$ employing a nested sampling algorithm. \par
By analyzing a large catalog of GW events, we found that peculiar velocities might be a source of systematic errors, depending on the direction of the bulk motion of galaxies. This behavior, however, affects mostly low redshift events, and progressively vanishes when adding farther away mergers to the inference. Moreover, it can be due to limited solid angle coverage of the light cone adopted, and a full sky catalog may address the problem in the first place. \par
We then performed \num{30} realizations of the sample of observed SBHBs assuming either \num{4} or \num{10} years of LISA operations. By analyzing them, we found that multiband observations of ``above-gap'' SBHBs can provide a competitive measurement of the cosmological parameters $\cosmoset = \{h, \Om\}$. In fact, assuming \num{4} (\num{10}) years of observation, the Hubble constant is determined down to a \num{1.5}$\%$ (\num{1.1}$\%$) precision, while $\Om$ is measured at a  \num{26.2}$\%$ (\num{14.3}$\%$) level. In general, the two parameters are estimated to better than $\sim2\%$ and $\sim30\%$ respectively. \par
We found, however, a couple of realizations yielding biased solutions, favoring $h$ and $\Om$ values railing against the upper end of the prior range. We traced back the insurgence of those solutions to a combination of factors, including the form of the likelihood, the $h-\Om$ degeneracy and the lack of very informative events in those realizations. We notice here that this problem did not appear in the work of \cite{del2018stellar} and \cite{laghi2021gravitational}, who employed the same techniques. This is likely because \cite{del2018stellar} considered ``below-gap'' SBHBs at $z<0.1$; at such low redshifts, the number of galaxies in the error volume is generally much smaller, and it is unlikely that all the events have support around $z_{\rm max}$. On the other hand, \cite{laghi2021gravitational} investigated a limited number of realizations of their experiment, perhaps insufficient to identify a bad one. By individually analyzing those ``bad realization'' we provided practical ways to recognize and discard spurious solutions, allowing the recovery of the correct cosmological parameters even in their presence. \par
Nevertheless, the identification of this issue calls for future improvements. Among them, we underline the need to upgrade the design of the simulation so that the binaries from our population merge in crowded clusters in the very first place, as reality suggests. This can be achieved by consistently assigning a host to each GW event before the Fisher matrix pipeline, thus avoiding the need of an artificial true host selection. We plan to enhance hosting probability assignments, so to take into account also for other important parameters besides the host candidate's sky position (e.g.,~the mass of the galaxy). Furthermore, a reweight or change of the shape of the prior can be folded into the analysis, ensuring that an uninformative experiment returns a flat posterior on the cosmological parameters. We defer a detailed investigation of these possible improvements of the analysis to future work. \par
Finally, our work relies on the assumption that SBHBs above the mass gap come only from the isolated evolution channel. However BHs inside and above the mass gap can be formed in dense environment thanks to the close interplay between dynamics and stellar evolution \cite{2020ApJ...903...45K, 2020MNRAS.497.1043D}. These additional sources would increase the number of detected systems that could be exploited in our approach, allowing us to constrain even further the cosmological parameters.

\section*{Acknowledgments}
A.S. acknowledges financial support provided under the European Union’s H2020 ERC Consolidator Grant ``Binary Massive Black Hole Astrophysics'' (B Massive, Grant Agreement No.: 818691).

\appendix
\section{GALAXY MASS THRESHOLD}\label{gal_mass_threshold}

In this study, we considered only galaxies with stellar mass $M_{\rm gal} \ge \SI{3e10}{\msun}$. The motivation behind this choice is two-fold. On one hand,  the vast majority of stars - and compact remnants - in the Universe is hosted in galaxies above this mass threshold; on the other hand, simulating large light cones down to a much lower mass threshold requires significant computing time and memory resources. It is, however, important to investigate to what extent our results are dependent on this specific choice. \par
\begin{figure}[b]
    \centering
    \includegraphics[width=0.5\textwidth]{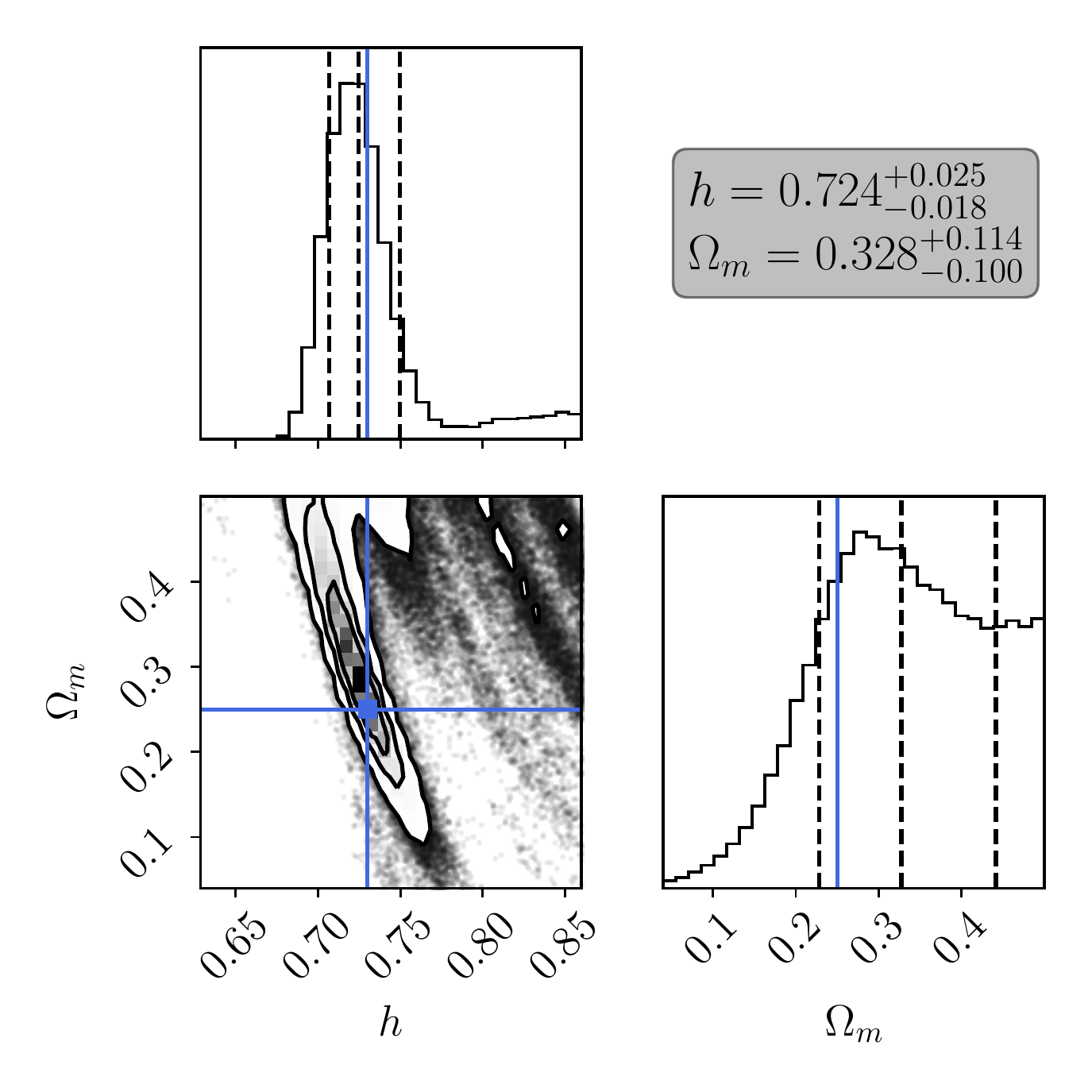}
    \caption{Averaged joint posterior distribution of \num{20} different realizations assuming a lower threshold in galaxy's selection.}
    \label{fig:lower_threshold_results}
\end{figure}

To this end, we ran a test light cone considering all galaxies with $M_{\rm gal} \ge \SI{e10}{\msun}$. We built the error boxes with the same procedure outlined in the main text, constructing \num{20} independent realizations of the above-gap SBHB population. The main result of this exercise is shown in \cref{fig:lower_threshold_results}. We notice here that, although the results are consistent with those obtained with a higher mass cut, the precision in the determination of the cosmological parameter is slightly degraded. In particular, the width of the marginalized posterior on $h$ is about $30\%$ larger, with an average error on its determination increasing from $\approx 1.5\%$ to $\gtrsim 2\%$. Similarly, constraints on  $\Om$ are slightly looser, with typical precision of $\approx 30\%$. The reason of this precision loss might lurk in the peculiar clustering properties of different galaxy populations. For example, specific classes of dwarf galaxies tend to be less clustered and more evenly distributed in the sky \citep{2021ApJS..252...18T}, which is expected to somewhat reduce the effectiveness of our methodology. Despite this, we notice that the main results of our study are robust and we defer a more comprehensive investigation of effects such as catalog completeness and selection effects to future work.

\newpage
\bibliography{bibliography.bib}

\end{document}